\documentclass[conference]{IEEEtran}
\IEEEoverridecommandlockouts

\usepackage{graphicx}
\ifCLASSOPTIONcompsoc
    \usepackage[caption=false, font=normalsize, labelfont=sf, textfont=sf]{subfig}
\else
\usepackage[caption=false, font=footnotesize]{subfig}
\fi
\usepackage{cite}
\usepackage{amsmath,amssymb,amsfonts}
\usepackage{comment}
\usepackage{cleveref}
\usepackage{listings}
\usepackage{multirow}
\usepackage[utf8]{inputenc}
\usepackage{algorithmic}
\usepackage{graphicx}
\usepackage{textcomp}
\usepackage{xcolor}
\def\BibTeX{{\rm B\kern-.05em{\sc i\kern-.025em b}\kern-.08em
    T\kern-.1667em\lower.7ex\hbox{E}\kern-.125emX}}
    
\begin{document}

\newcommand{\ourtool}{Vyasa}
\newcommand{\betterparagraph}[1]{\noindent\textbf{#1. }}

    \makeatletter
    \newcommand{\linebreakand}{%
      \end{@IEEEauthorhalign}
      \hfill\mbox{}\par
      \mbox{}\hfill\begin{@IEEEauthorhalign}
    }
    \makeatother
    
 \author{\IEEEauthorblockN{Prasanth Chatarasi}
    \IEEEauthorblockA{\textit{Georgia Tech} \\
    Atlanta, USA \\
    cprasanth@gatech.edu}
    \and
    \IEEEauthorblockN{Stephen Neuendorffer}
    \IEEEauthorblockA{\textit{Xilinx Research Labs} \\
    San Jose, USA \\
    stephenn@xilinx.com}
    \and
    \IEEEauthorblockN{Samuel Bayliss}
    \IEEEauthorblockA{\textit{Xilinx Research Labs} \\
    San Jose, USA \\
    samuelb@xilinx.com}
    \linebreakand 
    \IEEEauthorblockN{Kees Vissers}
    \IEEEauthorblockA{\textit{Xilinx Research Labs} \\
    San Jose, USA \\
    keesv@xilinx.com}
    \and
    \IEEEauthorblockN{Vivek Sarkar}
    \IEEEauthorblockA{\textit{Georgia Tech} \\
    Atlanta, USA \\
    vsarkar@gatech.edu}
    }
    
\title{Vyasa: A High-Performance Vectorizing Compiler for Tensor Convolutions on the Xilinx AI Engine}
\thispagestyle{plain}
\pagestyle{plain}

\maketitle

\begin{abstract}

Xilinx’s AI Engine is a recent industry example of energy-efficient vector processing that includes novel support for 2D SIMD datapaths and  shuffle interconnection network. The current approach to programming the AI Engine relies on a C/C++ API for vector intrinsics. While an advance over assembly-level programming, it requires the programmer to specify a number of low-level operations based on detailed knowledge of the hardware.  To address these challenges, we introduce {\em Vyasa}, a new programming system that extends the Halide DSL compiler to automatically generate code for the AI Engine.  We evaluated Vyasa on 36 CONV2D and 6 CONV3D workloads, and achieved geometric means of 7.6 and 23.3 MACs/cycle for 32-bit and 16-bit operands (which represent 95.9\% and 72.8\% of the peak performance respectively).  For 4 of these workloads for which expert-written codes were available to us, Vyasa demonstrated a geometric mean performance improvement of 1.10$\times$ with 50$\times$ smaller code relative to the expert-written codes.

\end{abstract}


\newsavebox{\firstlisting}
\begin{lrbox}{\firstlisting}
\begin{lstlisting}[language=C,escapeinside={@}{@}]
Buffer<int16> img_in(image_width, image_height);
Buffer<int16> kernel_coeff(3,3);
RDom r(3, 3);
Func img_out;
img_out(x,y) +=img_in(x+r.x,y+r.y)*kernel_coeff(r.x,r.y);
\end{lstlisting}
\end{lrbox}

\newsavebox{\secondlisting}
\begin{lrbox}{\secondlisting}
\begin{lstlisting}[language=C,escapeinside={@}{@}]
int16_t **  img_in,  kernel_coeff, img_out;
for(int x=0; x < image_height; x++)
 for(int y=0; y < image_width; y++)
  for(int r=0; r < 3; r++)
   for(int s=0; s < 3; s++)
    img_out[x][y] += img_in[x+r][y+s]*kernel_coeff[r][s];
\end{lstlisting}
\end{lrbox}

    

\section{Introduction}\label{sec:introduction}



It is widely recognized that a major disruption is under way in
computer hardware as processors strive to extend, and go beyond, the
end-game of Moore's Law.  
Unlike previous generations of hardware evolution, these ``extreme heterogeneity'' systems will have a profound impact on future
software.  As part of these trends, there is a strong resurgence of interest in improving vector processing (SIMD) units due to the significant energy efficiency benefits of using SIMD parallelism. 
These benefits increase with widening SIMD vectors, reaching vector register lengths of 2048 bits in the scalable vector extension of the Armv8 architecture~\cite{10.1109/MM.2017.35}.
Furthermore, there is an emphasis on specializing SIMD units to further improve energy efficiency benefits for specific domains such as Machine learning, Computer Vision, and 5G Wireless. 
An important specialization, which is referred to as ``2D vector SIMD datapath''~\cite{4205131,6113792,10.1145/1555815.1555773}, is the ability of each vector lane to execute more than one scalar operation and to chain the results from one operation to another.
Another specialization includes the removal of expensive data permutation units (e.g., shuffle units)~\cite{10.1145/1995896.1995938,10.5555/1788374.1788385} and instead introduce sophisticated, programmable interconnection  networks (a.k.a shuffle networks) between the SIMD datapath and vector register file to support the required data permutation patterns~\cite{10.5555/1812707.1812738,10.1145/1555815.1555773}.

A recent industry example with these specializations is the Xilinx Versal AI Engine~\cite{xilinx:aiengine}, a high-performance VLIW SIMD core 
which can deliver performance comparable to traditional FPGA solutions for Computer Vision, Deep Learning, and 5G wireless domains, but with 50\% less power consumption and up to eight times more compute capacity per silicon area~\cite{xilinx:aiengine}.
AI Engine cores are tightly integrated with programmable logic in Xilinx Versal ACAP devices to form a seamless heterogeneous compute platform~\cite{gaide2019xilinx,xilinx:versal} applicable to a wide variety of HPC applications.
Furthermore, the Versal AI Engine series VC1902 has a total of 400 AI Engines that together delivers a peak performance of 6.4 TOPS, 25.6 TOPS and 102.4 TOPS for 32-bit, 16-bit, and 8-bit operands, respectively~\cite{xilinx:versal}.

Tensor convolution is a widely used mathematical operation in these domains, and it is becoming increasingly  important  with the rise of its use in image processing workflows~\cite{ragan2013halide,conf/asplos/MullapudiVB15,pu2017programming} and with the proliferation of deep learning models~\cite{russakovsky2015imagenet,karpathy2015deep,toshev2014deeppose,farabet2013learning} in data centers, edge, and mobile devices.
There has been a lot of prior work in optimizing  tensor convolutions for a variety of target hardware devices such as CPUs~\cite{ragan2013halide,conf/asplos/MullapudiVB15,10.5555/3291168.3291211}, GPUs~\cite{ragan2013halide,journals/corr/ChetlurWVCTCS14,10.5555/3291168.3291211}, FPGAs~\cite{zhang2015optimizing,ma2017optimizing,zhao2019mRNA,zhang2018caffeine,10.5555/3291168.3291211}, and Dataflow accelerators~\cite{zhao2019mRNA,DBLP:journals/corr/abs-2002-07752,cyphers2018intel,jouppi2017datacenter}.
However, even for well understood applications like convolution, generating the best code for new high performance processor architectures from high-level descriptions can be challenging.  This work demonstrates the ability to automatically optimize tensor convolutions for the AI Engine and to obtain close to the peak performance for various workloads while using a high-level programming model, rather than low-level C/C++ intrinsics.

\betterparagraph{Challenges} Achieving peak performance on the AI Engine requires leveraging several architectural features to maximize vector datapath occupancy during program execution.
Unlike standard SIMD architectures which operate on 1D vectors, the AI Engine architecture includes 2D vector operations for some datatypes which conceptually implement the fusion of several 1D vector operations.  Also unlike other architectures, the AI Engine doesn't implement direct support for unaligned loads, scalar broadcasts, and data manipulation operations.  Instead, the AI Engine architecture includes a novel shuffle network which selects desired elements of a vector register for a vector operation instead of explicitly shuffling and storing them into another vector register.  In order to effectively leverage these features, the layout of data in memory must match the capabilities of the shuffle network.

Existing AI Engine compilers do not perform auto-vectorization, leaving it to expert programmers to explicitly write high-performance vector code using architectural intrinsic functions.  Optimizing programs in this way can be time-consuming even for experts.  At the same time, there are a wide variety of tensor convolution operators in common use, for instance, deep neural networks may contain regular 2D convolutions, depth-wise convolutions, and point-wise convolutions.  Even within the same network, the shape of tensor data can vary radically between the early and late layers in DNN models.  We find that no single optimization strategy is an optimal choice for all these scenarios.  Reducing the need for manual optimization and quickly adapting to new tensor operations through automatic optimization avoids these problems.

With all these challenges, the overall goal of our work is {\em to automate the generation of high-performance vector code for tensor convolutions based on their variations and shapes, while exploiting the unique capabilities of the Xilinx AI Engine without requiring manual effort in development and tuning}.  Achieving this goal requires significant loop-level reuse analysis, code transformation, and data-layout transformation, along with optimized low-level code generation taking into account the shuffle network and memory optimizations such as vector register reuse (including partial reuse)~\cite{stock2014vectorization,6113836}. 

The main technical contributions of this paper are briefly described below:
\begin{itemize}
    \item We introduce a new domain-specific intermediate representation called {\em Triplet} to symbolically capture the loop body of a tensor convolution, and  to simplify analyses and transformations required to generate high-performance code for the AI Engine.
    \item We propose a novel multi-step compiler approach which includes analyses and transformations to 1) exploit the 2D SIMD datapath by identifying multiple 1D logical vector operations that can be legally fused, 2) realize unaligned loads, scalar broadcasts, data manipulation using the shuffle network, 3) improve memory utilization by performing vector register reuse and also loop optimizations, and 4) generate  code that is more amenable to enabling  VLIW instruction scheduling for the AI Engine.
    \item We created a new tool, \ourtool{}\footnote{Vyasa means ``compiler'' in the Sanskrit language, and also refers to the sage who first compiled the Mahabharata.}, to implement our multi-step compiler approach.  \ourtool{} is built on the Halide framework~\cite{ragan2013halide} and includes extensions needed for the AI Engine that are not supported by Halide. Given a tensor convolution specification in the Halide language and workload sizes, \ourtool{} generates high-performance C-code with vector intrinsics for the AI Engine.
    \item 
    We evaluated Vyasa on 36 CONV2D and 6 CONV3D workloads using a cycle-accurate simulator\footnote{Since the AI Engine architecture was developed for real-time processing applications which require deterministic performance, the simulator results are reliably correlated with actual performance of the AI Engine hardware.}.
   Our results show geometric means of 7.6 and 23.3 MACs/cycle for 32-bit and 16-bit operands (which represent 95.9\% and 72.8\% of the peak performance respectively). For four of these workloads for which expert-written implementations were available to us, Vyasa achieved a geometric mean performance improvement of 1.10$\times$ from Halide code that is around 50$\times$ smaller than the expert-written C/C++ code.
    
\end{itemize}

\section{Background}
\label{sec:background}

In this section, we start with a brief overview of tensor convolutions, and then we briefly summarize the key architectural features of the Xilinx Versal AI Engine.
%

\subsection{Tensor Convolutions}
\label{subec:background_convolution}
A convolution is a mathematical operation which computes the amount of overlap of a function $g$ as it is shifted over another function $f$, and it is symbolically represented as $f \circ g$.
In this section, we restrict our attention to describing CONV2D, a popular convolution operator widely used in Deep learning~\cite{Alexnet,VGGnet,russakovsky2015imagenet,karpathy2015deep,toshev2014deeppose,farabet2013learning} and Computer Vision~\cite{shapiro:2001,ragan2013halide,conf/asplos/MullapudiVB15,pu2017programming}.
In these domains, the function {\tt f} and {\tt g} are referred to as the ``input'' tensor (a.k.a image/activations) and ``weight'' tensor (a.k.a filters/kernels), respectively.
The CONV2D deals with three four-dimensional tensors, i.e., Output (O), Weight (W), and Input (I), whose dimensions are described below.
\begin{table}[!ht]
\centering
\begin{tabular}{|c|c|c|c|c|}
\hline
\textbf{Tensor}     & \textbf{Dim1} & \textbf{Dim2} & \textbf{Dim3} & \textbf{Dim4} \\ \hline
\textbf{Output (O)} & Width (X)     & Height (Y)    & Channels (K)  & Batch (N)     \\ \hline
\textbf{Weight (W)} & Width (R)     & Height (S)    & Channels (C)  & Batch (K)     \\ \hline
\textbf{Input (I)}  & Width (X')    & Height (Y')   & Channels (C)  & Batch (N)     \\ \hline
\end{tabular}
\end{table}

The mathematical expression of the CONV2D operations is shown below, where $f$ refers to stride factor.
\begin{align*}
O(x,\mbox{ }  y,\mbox{ }  k,\mbox{ }  n) = & \sum_{c}^{C}  \sum_{s}^{S} \sum_{r}^{R}  W(r,\mbox{ } s,\mbox{ } c,\mbox{ } k) \\
& \times I(x \times f + r, \mbox{ } y \times f + s,\mbox{ }  c,\mbox{ }  n)
\end{align*}

The convolutions used in Computer Vision are special cases of the CONV2D operator, where each tensor has only the first two dimensions (width and height) and stride factor set to one. However, there exist a wide variety of filter sizes (ranging from 2 to 11) used in many different image processing operators, such as Gaussian smoothing and edge detection~\cite{shapiro:2001}.

A wide variety of other specialized variations of the CONV2D operator are used in Convolutional Neural Networks such as point-wise, depth-wise separable, and spatially separable convolutions. 
These variations can be viewed as constraints on the regular CONV2D operator, and are shown below.

\begin{table}[!ht]
\centering
\begin{tabular}{|c|c|}
\hline
\textbf{Operator}             & \textbf{Constraints on CONV2D}                                                                    \\ \hline
\textbf{Point-wise (PW)}           & Filter width = Filter height = 1                                                                  \\ \hline
\textbf{Fully-connected (FC)}      & \begin{tabular}[c]{@{}c@{}}Filter width = Input width\\ Filter height = Input height\end{tabular} \\ \hline
\textbf{Spatially separable (SS)}  & Filter width = 1 or Filter height = 1                                                             \\ \hline
\textbf{Depth-wise separable (DS)} & Input channels = Filter channels = 1                                                              \\ \hline
\end{tabular}
\end{table}

Even though we briefly described the CONV2D operator and its variations, our approach is applicable to other convolution operators such as CONV1D and CONV3D.

\subsection{Xilinx AI Engine}
Driven by the performance and energy efficiency requirements of many computing applications, Xilinx introduced Versal Advanced Compute Acceleration Platform (ACAP)~\cite{gaide2019xilinx,xilinx:versal}, a fully software-programmable, heterogeneous compute platform.
The Versal platform consists of three types of programmable processors -- Scalar Engines (CPUs), Adaptable Engines (Programmable Logic), and an array of Intelligent Engines (AI Engines)~\cite{xilinx:versal}.
In this work, we focus on AI Engines, which are specialized SIMD and VLIW high-performance processors for compute-intensive applications such as computer vision, machine learning workloads, and 5G wireless. 
AI Engines are highly energy efficient compared to FPGAs and can deliver up to 8X silicon compute density at 50\% the power consumption of traditional FPGA solutions~\cite{xilinx:aiengine}. 

An AI Engine includes a 2D SIMD datapath for fixed-point vector operations (our focus), a 1D SIMD datapath for floating-point vector operations, and a scalar unit for scalar operations.
Each AI Engine also has access to 128KB scratchpad (a.k.a data/local) memory, a 16KB program memory, and a 256B vector register file (a total of 16 registers with each size being 128 bits). 
These high-performance AI Engines are programmed using the C/C++ programming language with optional pragmas. 
A simplified overview of the key architectural features of the AI Engine core is shown in~\cref{fig:AIEngine-Architecture-in-depth}, and these features are briefly described below.

\begin{figure}[!ht]
    \centering
    \includegraphics[scale=0.4]{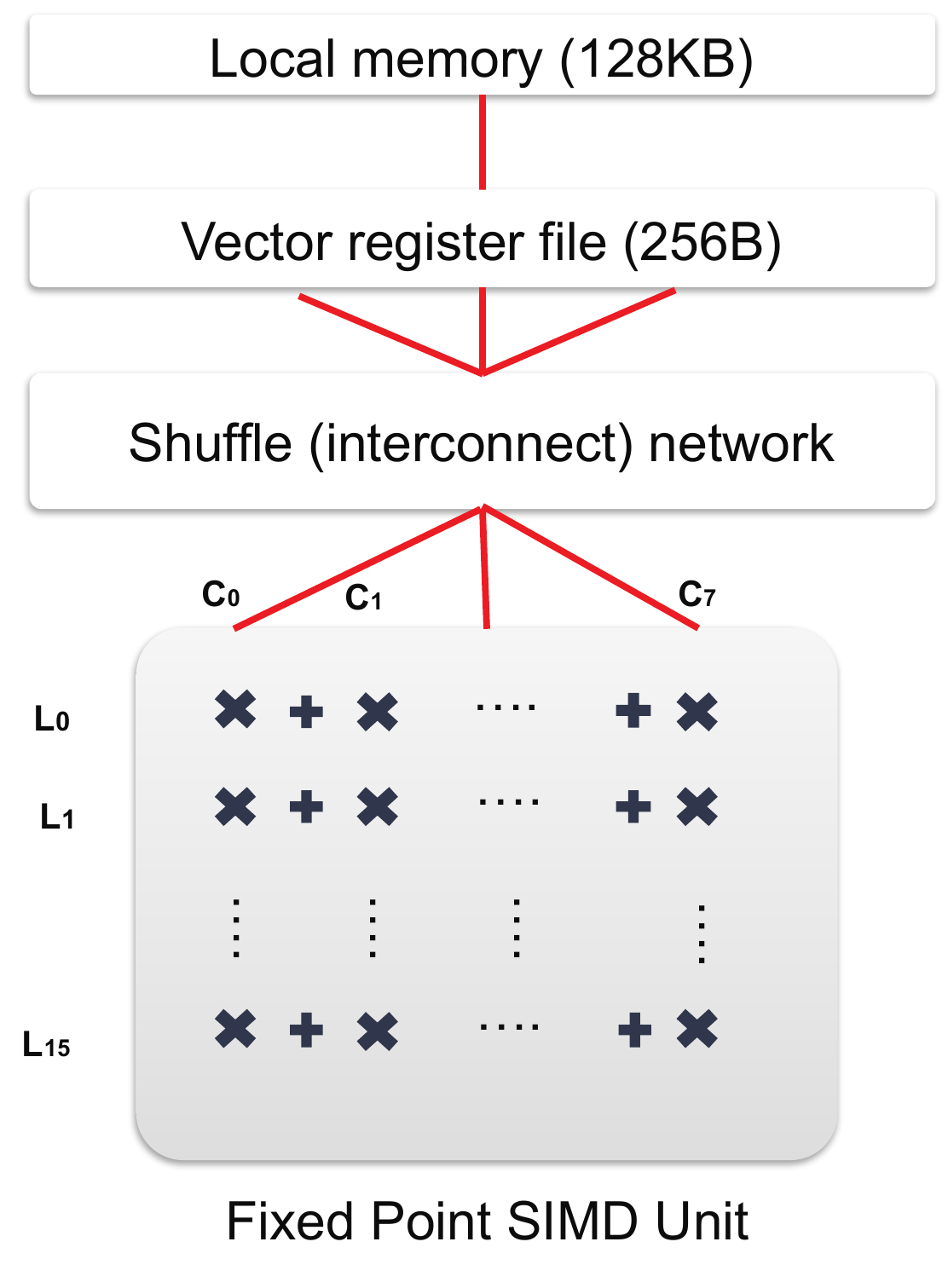}
    \caption{A pictorial overview of the key architectural features of the Xilinx AI Engine, i.e., 2D vector SIMD datapath and shuffle network.}
    \label{fig:AIEngine-Architecture-in-depth}
\end{figure}

\betterparagraph{1) Two-dimensional SIMD Datapath}
The fixed point vector unit of the AI Engine is a two-dimensional SIMD datapath, and vector operations on the 2D SIMD datapath are described using {\em lanes/rows} and {\em columns}.
The number of lanes corresponds to the number of output values generated from the vector operation.
The number of columns is the number of operations that are done per output lane, with each of the results being reduced together. 
This technique of executing back to back dependent scalar operations along a vector lane is popularly known as operation chaining~\cite{4205131} and can improve energy efficiency by not writing intermediate values back to the register file.
Furthermore, the number of columns is dependent on the operand precision. Operations on 32-bit types are organized as 8 lanes with 1 column, without internal reduction.  Operations on 16-bit types are organized as either 16 lanes with 2 columns or 8 lanes with 4 columns. Operations on 8-bit types are organized as 16 lanes with 8 columns.
As a result, the 2D datapath can perform either 8 MACs on 32-bit inputs, 32 MACs on 16-bit input, or 128 MACs on 8-bit input per cycle.


\betterparagraph{2) Shuffle network} A key novelty of the AI Engine architecture is its {\em shuffle network}, a flexible interconnection network between the 2D SIMD datapath and vector register file to allow flexible data selection from the input vector registers for the multipliers of each lane and column of the SIMD datapath.
The ability to configure the shuffle network for each vector operation is exposed to programmers via the arguments of the vector intrinsic functions.
Unlike the data manipulation units in traditional SIMD units, the data selection using the shuffle network over a vector register can only be used during a vector operation. 
The granularity of data selection using the shuffle network on the vector registers is 32b, and so the network allows full flexibility for making data selection, replication, and permutation on vectors of 32b data types.
However, for data types of smaller sizes such as 16b and 8b data types, the shuffle network imposes further constraints on data selection.


Vector loads and stores in the AI Engine must be aligned to 128-bit data memory boundaries.
The AI Engine does not implement unaligned loads or scalar broadcasts.
Instead, these operations are typically realized/implemented using a combination of aligned loads and configuration of the shuffle network.

\betterparagraph{3) VLIW capabilities} The AI Engine has support for very long instruction word (VLIW) that can provide up to 6-way instruction parallelism to hide long instruction latencies.
The VLIW instruction includes two scalar operations, two vector load operations, one vector store operation, and one fixed/floating-point vector operation.
The AI Engine compilers have support for automatic software pipelining~\cite{10.1145/53990.54022} of innermost loops to exploit instruction-level parallelism.
\section{Our Approach}\label{sec:our-approach}

In this section, we introduce our approach to generating high-performance vector code for a given high-level specification of tensor convolution and its workload sizes that fit into a single AI Engine's data memory. 
These vector codes are intended to execute on a single AI Engine and will be integrated by a high-level compiler to run larger tensor convolutions across multiple AI Engines.
Our approach is summarized in ~\cref{fig:workflow} and is implemented in a tool called \ourtool{}.
The tool is developed as an extension to the Halide framework~\cite{ragan2013halide}.
\begin{figure}[!ht]
    \centering
    \includegraphics[width=\linewidth]{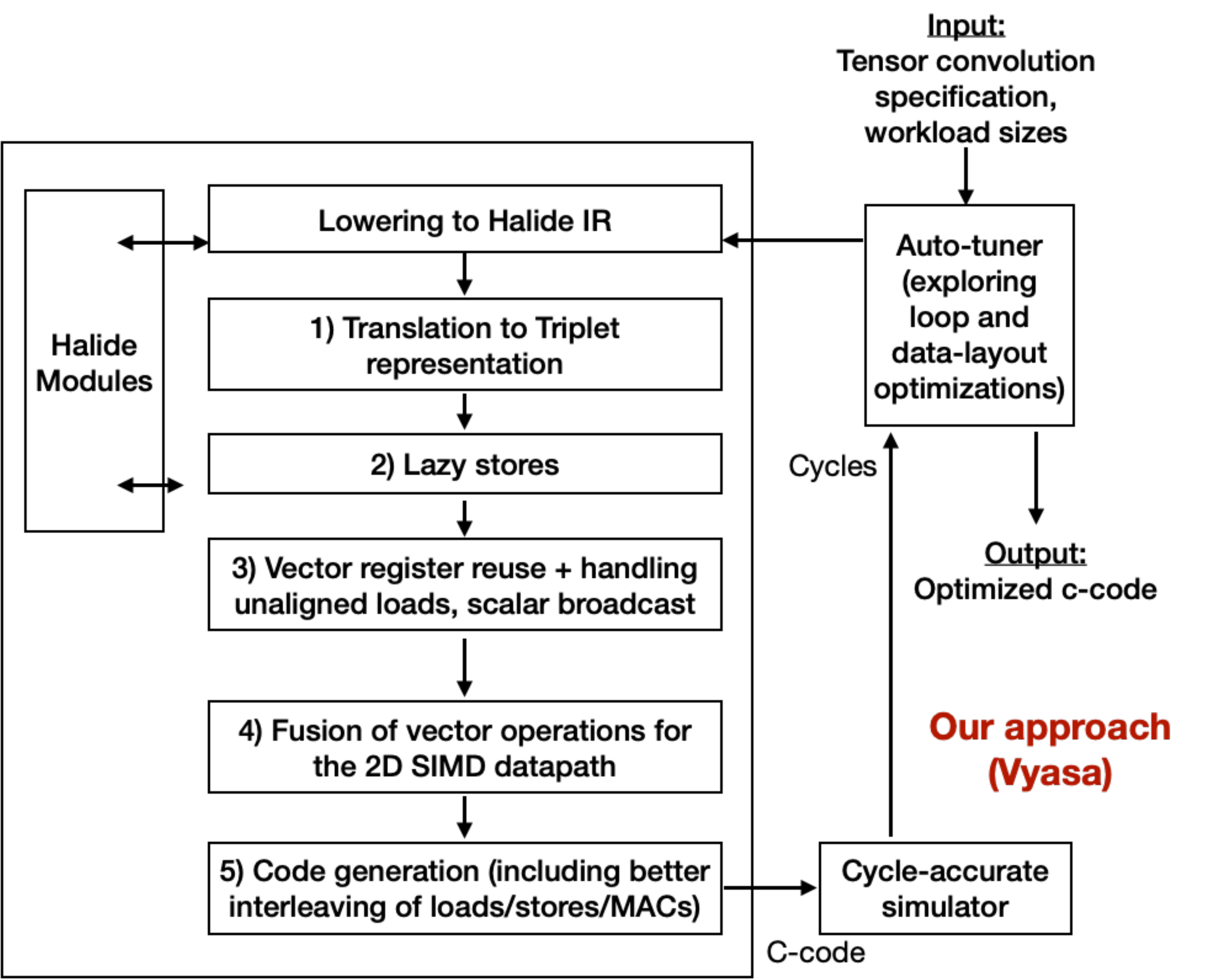}
    \caption{Workflow of our approach (\ourtool{}) which is implemented as an extension to the Halide framework~\cite{ragan2013halide}.}
    \label{fig:workflow}
\end{figure}
Our approach begins with an auto-tuner taking the specification of a tensor convolution in the Halide language and also the corresponding workload sizes.
Then, the auto-tuner iterates through each possible schedule in the space of loop transformations and data-layouts, and invokes our multi-step compiler approach to generate high-performance vector c-code corresponding to the schedule.
Then, our approach evaluates the generated code using a cycle-accurate simulation of the AI Engine, and chooses the best one among all schedules to finally emit as the performant output code.

Our multi-step compiler approach starts from the specification of a tensor convolution, a schedule from the auto-tuner, and workload sizes.
It consists of the following steps:
\begin{enumerate}
    \item Transforming the loop body of the tensor convolution operation in the Halide IR (after lowering) into our symbolic {\em triplet} representation for convenience in doing analyses, transformations, and code generation,
    \item Performing 'lazy stores' optimization by accumulating all partial (intermediate) results of an output before generating a store to reduce the memory traffic,
    \item Exploiting vector register reuse, and realizing unaligned loads and scalar broadcast operations using the shuffle (interconnection) network, 
    \item Identifying suitable 1D logical vector operations (multiplications) that contribute to same output through accumulation/reduction and fusing them into operations matching the 2D SIMD datapath,
    \item Interleaving load and store operations with vector operations to make it easy for the AI Engine compilers to perform VLIW instruction scheduling,
    \item Generating C-code with vector intrinsics.
\end{enumerate}

%


\subsection{Translating into Triplet Representation}\label{subsec:converting-to-triplet-data-structure}

\begin{figure}[!htb]
\centering
\begin{lstlisting}[language=C,escapeinside={@}{@}]
Buffer<int16> I(X',Y'); Buffer<int16> W(4,3);
Var x, y; RDom r(4, 3); Func O; //output

//(a) Description of the convolution computation
O(x,y) += W(r.x, r.y) * I(x+r.x, y+r.y);

//(b) A sample schedule: Unrolling reduction loops 
//Vectorizing loop corresponding to image width
O.update().unroll(r.x, 4).unroll(r.y,3)
     .vectorize(x, 16);

//(c) Intermediate code after lowering
for y:
 for x: (vectorized)
   O(x:x+15,y) += W(0,0) * I(x:x+15,y);
   O(x:x+15,y) += W(1,0) * I(x+1:x+16,y);
   O(x:x+15,y) += W(2,0) * I(x+2:x+17,y);
   O(x:x+15,y) += W(3,0) * I(x+3:x+18,y);
   ......
\end{lstlisting}
\caption{Algorithmic description of the convolution of a 4x3 filter  over an input 2D image in the Halide language~\cite{ragan2013halide}. A(a:b,c) is a short hand vector notation for denoting a contiguous slice from A(a,c) to A(b,c) in one direction.}
    \label{fig:motivating-example}
\end{figure}
In general, tensor convolutions are specified/implemented as multi-dimensional perfectly nested loops, where each statement of the loop body 
has two  aspects -- 1)  A group of multiply-and-accumulate (MAC) operations over input and weight tensors, and 
2) An update (reduction) operation to the output tensor
%
Since each statement in the convolution loop body performs a reduction operation and the reduction is commutative, the order of each statements doesn't impact its correctness. 
Hence, a representation holding information for each statement about the two major aspects described above is sufficient to capture the body precisely.
We call this representation a ``triplet'' since it holds information about the access patterns of the two operands of each multiplication and the update operand of each statement symbolically.
%

\begin{figure}[!ht]
    \centering
    \includegraphics[width=\linewidth]{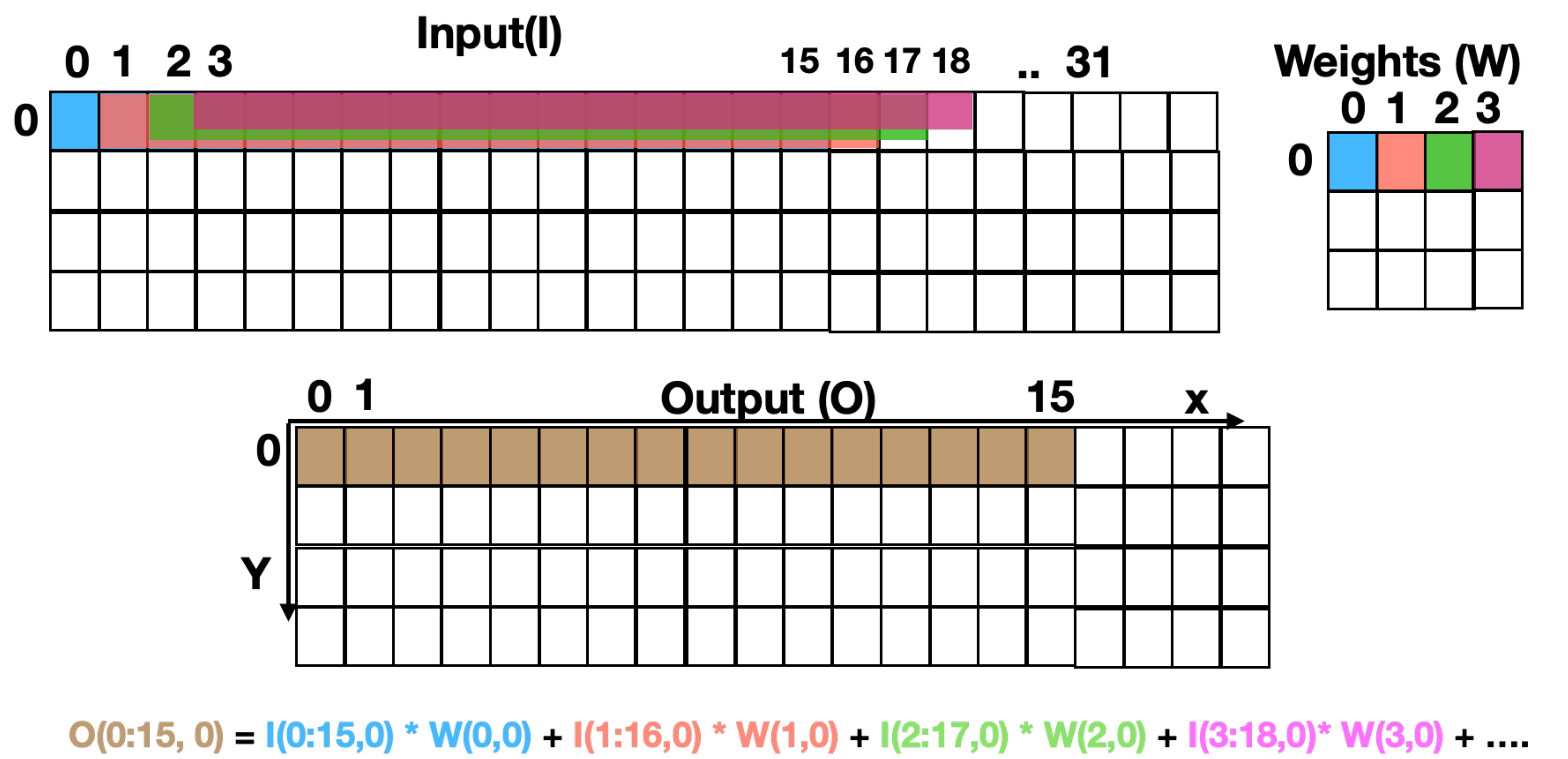}
    \caption{A pictorial overview of the convolution of 4x3 filter based on the schedule described  in~\cref{fig:motivating-example}(b) at the loop iterations x = 0 and y = 0.}
    \label{fig:approach-schedule}
\end{figure}

We consider the convolution of a filter with size 4x3
on an input with size {\tt X'} x {\tt Y'} as a running example (shown in~\cref{fig:motivating-example}(a)) to illustrate each step of our compiler approach.
A sample schedule for the above convolution is shown in~\cref{fig:motivating-example}(b), which refers to unrolling loops corresponding to filter dimensions ({\tt r.x, r.y}) and vectorizing the {\tt loop-x} with vector length as 16.
A pictorial overview of the computation at the loop iterations x = 0, y = 0 is shown in~\cref{fig:approach-schedule}.

After lowering the convolution specification using the schedule into the Halide IR, the first step in our approach is to translate the convolution loop body into our triplet representation.
For instance, the triplet representation of the loop body in~\cref{fig:motivating-example}(c) is shown in Table~\ref{tab:motivating-example-triplet-data-structure}, where each row in the table symbolically captures the access patterns of multiplication operands and update operands of a statement in the loop body. 

\begin{table}[!ht]
\centering
\caption{Triplet representation of the loop body in~\cref{fig:motivating-example}(c)}
\scalebox{1}{
\begin{tabular}{|c|c|c|}
\hline
\multirow{2}{*}{\textbf{\begin{tabular}[c]{@{}c@{}}Update Operation\\ Operand \end{tabular}}} & \multicolumn{2}{c|}{\textbf{MAC Operations}} \\ \cline{2-3} 
 & \textbf{\begin{tabular}[c]{@{}c@{}}Operand1\end{tabular}} & \textbf{\begin{tabular}[c]{@{}c@{}}Operand2\end{tabular}} \\ \hline
{\tt O(x:x+15, y)} & {\tt W(0, 0)} & {\tt I(x:x+15, y)} \\ \hline
{\tt O(x:x+15, y)} & {\tt W(1, 0)} & {\tt I(x+1:x+16, y)} \\ \hline
{\tt O(x:x+15, y)} & {\tt W(2, 0)} & {\tt I(x+2:x+17, y)} \\ \hline
{\tt O(x:x+15, y)} & {\tt W(3, 0)} & {\tt I(x+3:x+18, y)} \\ \hline
.. & .. & .. \\ \hline
\end{tabular}
}

\label{tab:motivating-example-triplet-data-structure}
\end{table}

\subsection{Lazy Stores Optimization } \label{subsec:lazy-writes}

An approach to code generation based on the triplet representation involves generating a vector store for each row of the representation.
But, this code generation can result in immediately writing multiplication results to the data memory causing more traffic. 
We introduce ``lazy stores'' optimization to delay writing the multiplication results of an output until there are no operations that can contribute to output.
The optimization works by grouping all the rows of the triplet representation contributing to the same output.
The benefits of the optimization can be observed in the presence of multiple statements in the loop body contributing to the same output.
An example of such behavior is seen in Table~\ref{tab:motivating-example-triplet-data-structure}, where all the statements contribute to the same output ({\tt O(x:x+15,y)}), and all these statements can be grouped into a single group (Table~\ref{tab:motivating-example-triplet-data-structure-after-lazy-writes}).
Now, the code generation involves generating a single vector store for each group, instead of generating for each row of the triplet representation.



\begin{table}[!htb]
\caption{Triplet representation after the lazy stores optimization}
\centering
\scalebox{1}{
\begin{tabular}{|c|c|c|}
\hline
\multirow{2}{*}{\textbf{\begin{tabular}[c]{@{}c@{}}Update Operation\\ Operand\end{tabular}}} & \multicolumn{2}{c|}{\textbf{MAC Operations}} \\ \cline{2-3} 
 & \textbf{\begin{tabular}[c]{@{}c@{}}Operand1\end{tabular}} & \textbf{\begin{tabular}[c]{@{}c@{}}Operand2\end{tabular}} \\ \hline
\multirow{4}{*}{{\tt O(x:x+15, y)}} & {\tt W(0, 0)} & {\tt I(x:x+15, y)} \\ \cline{2-3} 
 & {\tt W(1, 0)} & {\tt I(x+1:x+16, y)} \\ \cline{2-3} 
 & {\tt W(2, 0)} & {\tt I(x+2:x+17, y)} \\ \cline{2-3} 
 & {\tt W(3, 0)} & {\tt I(x+3:x+18, y)} \\ \cline{2-3} 
 & .. & ..  \\ \hline
\end{tabular}
}

\label{tab:motivating-example-triplet-data-structure-after-lazy-writes}
\end{table}


\subsection{Exploiting Vector Register Reuse \& Realizing Unaligned Loads and Scalar Broadcast} \label{subsec:vector-reuse-optimization}

Our approach leverages the AI Engine architecture's unique shuffle network to realize unaligned vector loads and scalar broadcast operations which are common in vectorization of tensor convolutions.
Our approach further uses the network to exploit the vector register reuse opportunities.

\betterparagraph{1) Realizing unaligned vector loads}
Simple vectorization of tensor convolutions often result in unaligned vector loads.  For example, if the vector load {\tt (I(x:x+15,y))} in~\cref{fig:motivating-example} is aligned to the boundary, then the subsequent vector loads such as {\tt I(x+1:x+16,y)} are unaligned.
Prior work on vectorization for SIMD architectures having no unaligned load/store support address this by generating two adjacent aligned loads covering the required load and using data manipulation/shuffle (register-to-register) instructions to realize an unaligned vector load~\cite{6113836}. 
Since the AI Engine architecture doesn't support unaligned loads or shuffle instructions, an alternative solution is necessary. 
Our approach leverages the AI Engine architecture support for grouping vector registers into a larger vector register.
Then, our approach constructs a larger aligned vector load which subsumes the required unaligned load and selects the data corresponding to the original unaligned vector load using the shuffle network.
For instance, the unaligned vector load {\tt I(x+1:x+16,y)} can be realized through a larger aligned vector load {\tt I(x:x+31,y)} and appropriate data selection parameters during vector operations on the load.

\betterparagraph{2) Exploiting vector register reuse}
Tensor convolutions often exhibit significant data reuse between vector loads. 
For instance, the two vector loads {\tt I(x:x+15,y)} and {\tt I(x+1:x+16,y)} have 15 data elements in common.
Exploiting vector register reuse by reusing those common elements instead of fetching again from the data memory is important to reduce memory traffic and achieve better performance.
Our approach groups individual vector loads having such reuse and constructs a larger aligned vector load that subsumes the individual vector loads having reuse.
During the vector operations, the individual vector loads are realized through appropriate data selection on the larger vector using the shuffle network.

Our approach implements the above idea by constructing a {\em reuse graph}, an undirected graph where each node denotes a vector load in the triplet representation and an edge is constructed between two nodes if they have at least one common element between them, i.e., presence of a reuse.
The reuse graph corresponding to the vector loads of the tensor {\tt I} in~\cref{tab:motivating-example-triplet-data-structure-after-lazy-writes} is shown in~\cref{fig:vector-loads-reuse-graph}, for instance, nodes {\tt I(x:x+15,y)} and {\tt I(x+1:x+16,y)} corresponds to two vector loads and the edge between them denotes the presence of common elements/reuse.

\begin{figure}[!ht]
    \centering
    \includegraphics[scale=0.15]{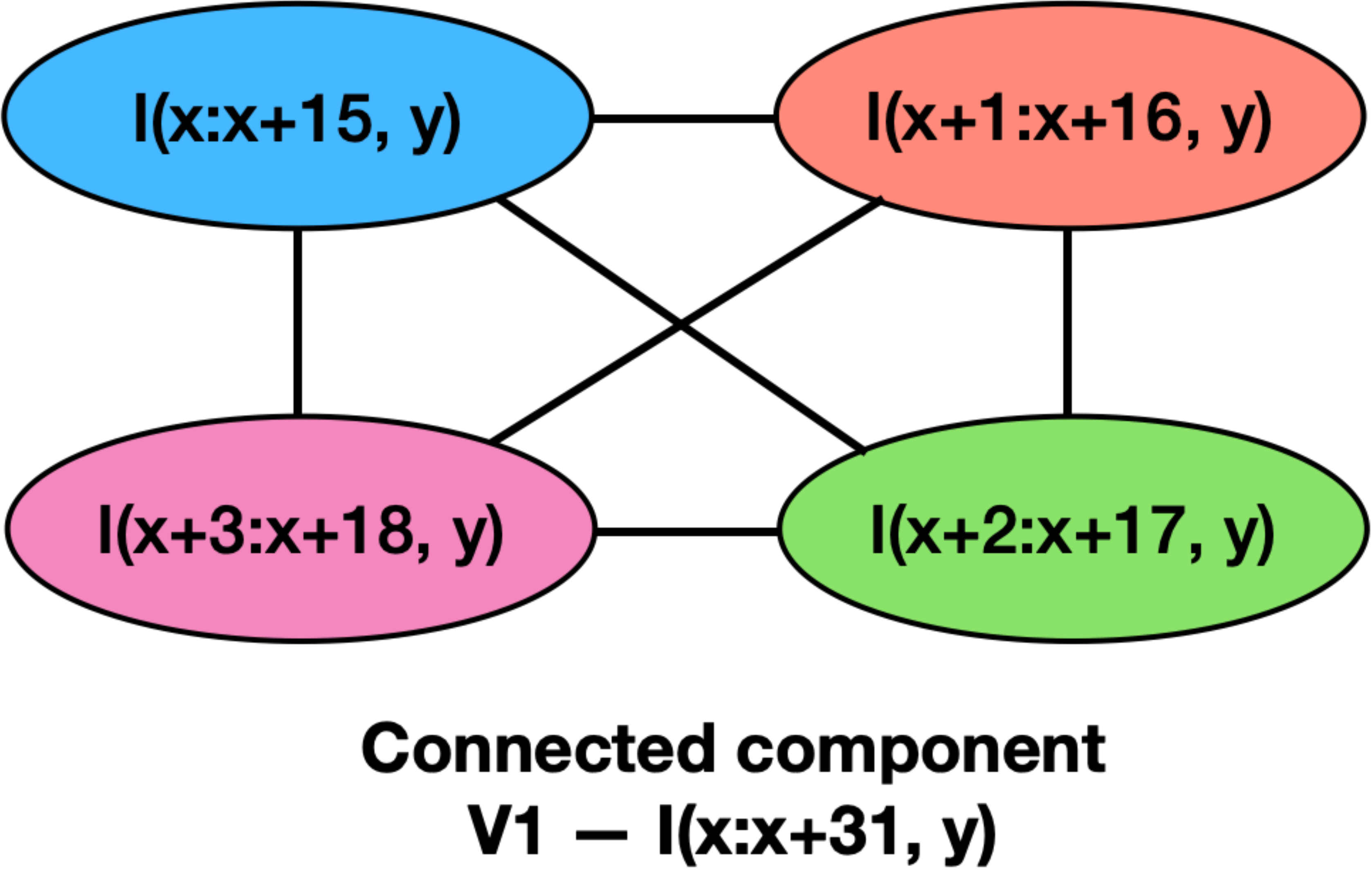}
    \caption{Reuse graph corresponding to the vector loads of the tensor {\tt I} in Table~\ref{tab:motivating-example-triplet-data-structure-after-lazy-writes}, and its connected components to construct larger vector loads.}
    \label{fig:vector-loads-reuse-graph}
\end{figure}

After constructing the reuse graph, our approach identifies connected components in the reuse graph, where each component represents a larger vector load that subsumes the individual vector loads in that component.
For instance, the connected component {\tt I(x:x+31,y)} in ~\cref{fig:vector-loads-reuse-graph} represents a larger vector load subsuming the vector loads {\tt I(x:x+15,y), I(x+1:x+16,y), I(x+2:x+17,y)}, and {\tt I(x+3:x+18,y)}.
Since our approach hasn't yet fused the logical 1D vector operations to exploit all columns of the 2D SIMD datapath, computing the data selection parameters is deferred to a later step (\cref{subsec:2dsimdunit}).
After replacing each individual vector load with its corresponding larger load, the running example results in having only three larger vector loads instead of twelve individual vector loads for the tensor {\tt I}.


\betterparagraph{3) Realizing scalar broadcasts} Similar to unaligned vector loads, vectorization of tensor convolutions involve scalar operands and require the support for scalar to vector broadcast operation, for, e.g., the scalar operand {\tt W(0,0)} in Table~\ref{tab:motivating-example-triplet-data-structure-after-lazy-writes}.
A naive approach to realize the broadcast operation of a scalar operand is by loading an aligned vector covering the operand and then using the shuffle network to select the the operand for all the lanes.
A downside of the above approach is that it may result in loading an entire vector while using only one value fetched from memory.

The scalar operands in the tensor convolutions typically exhibit significant spatial locality, e.g., the scalar operands such as {\tt W(0,0)} and {\tt W(1,0)} in Table~\ref{tab:motivating-example-triplet-data-structure-after-lazy-writes} are contiguous in the data memory. 
Similar to our approach in exploiting vector register reuse, we construct another reuse graph to identify scalar operands that are adjacent in data memory and can be subsumed as part of a single vector load. 
For instance, the operands {\tt W(0,0)}, {\tt W(1,0)}, {\tt (2,0)}, {\tt W(3,0)} can be realized over a vector load (say V2) of {\tt W(0:7, 0)}.

We represent the data selection of a set of values from a vector register using the shuffle network during a vector operation as {\em SELECT(V, \{$s_{ij}$\}))} where V represents the vector register and $s_{ij}$ denotes the index of the required element in the register V for the i$^{th}$ lane and j$^{th}$ column multiplier in the 2D SIMD datapath.
Our approach defers the computation of data selection parameters to the next step.
The triplet representation after realizing the unaligned loads, scalar broadcasts, and exploiting vector register reuse is shown in~\cref{tab:motivating-example-triplet-data-structure-after-scalar-to-vector}.

\begin{table}[!ht]
\centering
\caption{Triplet representation after addressing unaligned loads, scalar broadcast, and exploiting vector register reuse.}
\scalebox{1}{
\begin{tabular}{|c|c|c|}
\hline
\multirow{2}{*}{\textbf{\begin{tabular}[c]{@{}c@{}}Update Operation\\ Operand \end{tabular}}} & \multicolumn{2}{c|}{\textbf{MAC Operations}} \\ \cline{2-3} 
 & \textbf{\begin{tabular}[c]{@{}c@{}}Operand1\end{tabular}} & \textbf{\begin{tabular}[c]{@{}c@{}}Operand2\end{tabular}} \\ \hline
\multirow{4}{*}{{\tt O(x:x+15, y)}} & {\tt SELECT(V2, \{ \})} & {\tt SELECT(V1, \{ \})} \\ \cline{2-3} 
 & {\tt SELECT(V2, \{ \})} & {\tt SELECT(V1, \{ \})} \\ \cline{2-3} 
 & {\tt SELECT(V2, \{ \})} & {\tt SELECT(V1, \{ \})} \\ \cline{2-3} 
 & {\tt SELECT(V2, \{ \})} & {\tt SELECT(V1, \{ \})} \\ \cline{2-3} 
 & .. & .. \\  \hline
\end{tabular}
}
\label{tab:motivating-example-triplet-data-structure-after-scalar-to-vector}
\end{table}

\subsection{2D Vector SIMD Datapath} \label{subsec:2dsimdunit}

A key distinguishing feature of the AI Engine relative to the traditional SIMD units is the presence of a two-dimensional SIMD datapath which performs reduction across all columns of a SIMD lane.
A single 1D logical vector operation can occupy a single column of 2D datapath, but the vector operations on the 2D SIMD datapath require using all the columns of the datapath and don't allow partial utilization. 
Hence, our approach identifies and logically groups (fusing) all suitable 1D logical vector operations that contribute to the same output through accumulation/reduction and use the same set of vector register operands. 
The identification is done by searching in the triplet representation for operations having the same update operand and the same set of vector registers as multiplication operands.
Finally, our approach partitions the logical groups based on the number of columns available for the given operand type and also constraints imposed by the shuffle network on the data selection over vector register operands.
If the data selection required for the operands of fused vector operations is incompatible with the constraints of the shuffle network, then our approach generates a compilation error and prunes that candidate code variant.

\begin{figure}[!ht]
    \centering
    \includegraphics[width=\linewidth]{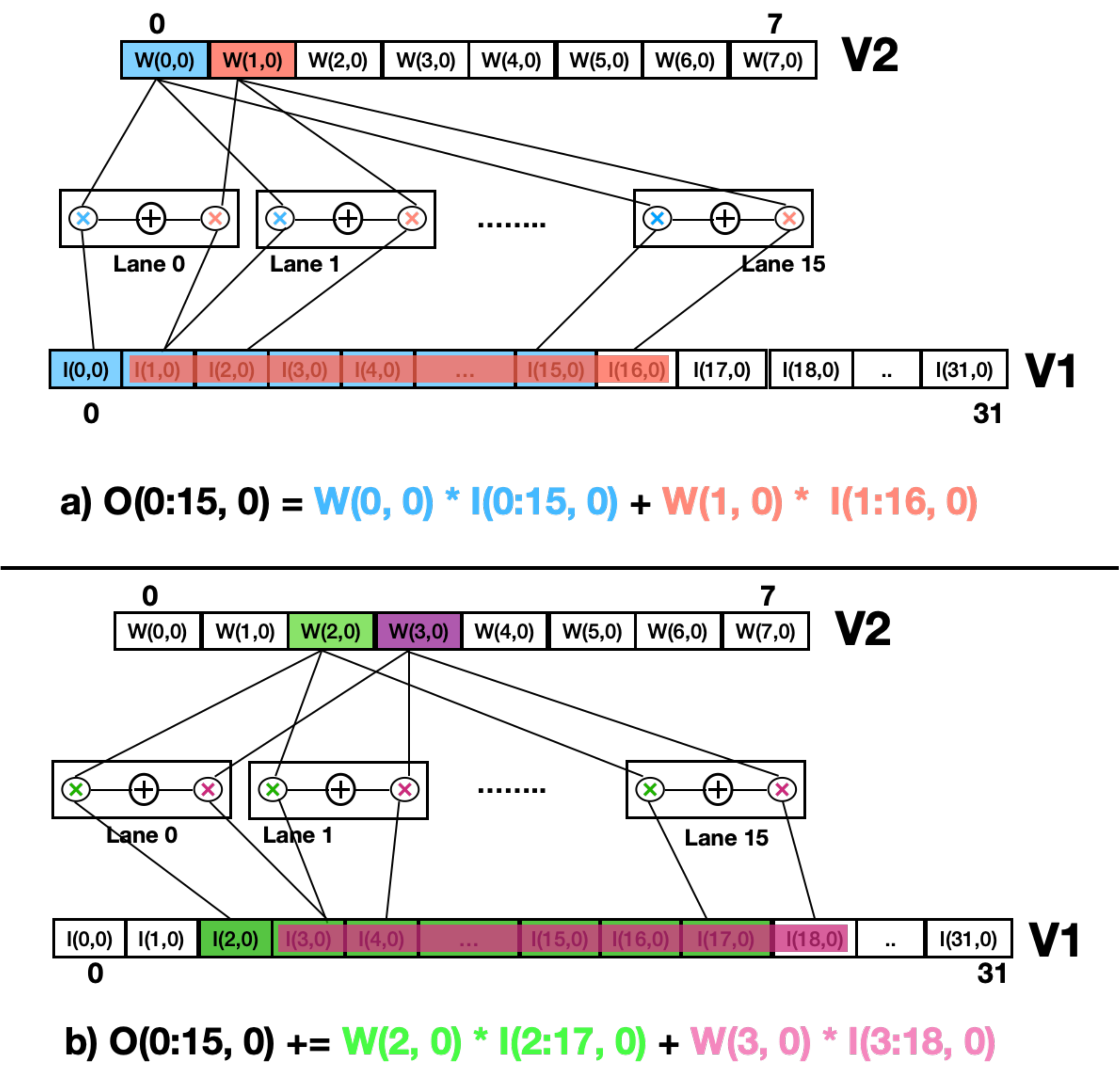}
    \caption{An overview of the two fused vector operations (a and b) over the vector registers V1, V2 for input and weights, respectively of the running example shown in Table~\ref{tab:motivating-example-triplet-data-structure-after-lazy-writes} at x=0 and y=0. The shuffle network of the AI Engine helps each multiplier of the 16 lanes and 2 columns of the 2D SIMD datapath to choose required elements from the vector registers.}
    \label{fig:example}
\end{figure}

There are four valid fusible logical 1D operations for each row of the filter in Table~\ref{tab:motivating-example-triplet-data-structure-after-lazy-writes}, our approach groups them into two fused vector operations whose overview is described in~\cref{fig:example}.
Furthermore, the triplet representation after fusing the logical 1D vector operations is shown in Table~\ref{tab:motivating-example-triplet-data-structure-after-fusing-macs}. 


\begin{table}[!ht]
\caption{Triplet representation after fusing the logical 1D vector multiplications and finding the data selection parameters}
\centering
\scalebox{0.95}{
\begin{tabular}{|c|c|c|}
\hline
\multirow{2}{*}{\textbf{\begin{tabular}[c]{@{}c@{}}Update operation\\ operand\end{tabular}}} & \multicolumn{2}{c|}{\textbf{MAC Operations (after fusing)}}                                                                                   \\ \cline{2-3} 
                                                                                                          & \textbf{\begin{tabular}[c]{@{}c@{}}Operand1\end{tabular}} & \textbf{\begin{tabular}[c]{@{}c@{}}Operand2\end{tabular}} \\ \hline
\multirow{3}{*}{{\tt O(x:x+15,y)}}                                                                              & {\tt SELECT(V2, \{j\})}                                                     & {\tt SELECT(V1, \{i+j\})}                                                   \\ \cline{2-3} 
                                                                                                          & {\tt SELECT(V2, \{j+2\})}                                                   & {\tt SELECT(V1, \{i+j+2\})}                                                 \\ \cline{2-3} 
                                                                                                          & ..                                                                    & ..                                                                    \\ \hline
\end{tabular}
}

\label{tab:motivating-example-triplet-data-structure-after-fusing-macs}
\end{table}

\subsection{Code Generation}\label{subsec:code-generator}

Our approach extends the code generation capabilities in the Halide~\cite{ragan2013halide} by implementing a code generator for the triplet representation to generate explicitly vectorized code using AI Engine intrinsic functions.
A naive approach to code generation can be implemented by first emitting all vector loads, followed by all vector MAC operation, and then finally all vector stores.
However, this naive approach results in variables (loads) having large live ranges, possibly leading to register spills and preventing software pipelining.  
Furthermore, optimization of memory accesses can be challenging for the downstream compilers only given the generated intrinsic code.
Hence, our approach reorders memory accesses and interleaves them with vector MAC operations during the code generation process to reduce the live range of each variable.
This process is relatively easy given the information about memory access patterns in Halide and helps the downstream compilers to improve packing of stores, loads, and vector MACs into VLIW instructions.
A snippet of the final code generated by our approach with interleaving of loads, vector operations, and stores over the running example is shown in~\cref{fig:code-generated}.

\begin{figure}[!htb]
\centering
\begin{lstlisting}[language=C,escapeinside={@}{@}]
//Generated code
for(int y=0; y < Y; y++)
 for(int x=0; x < X; x+=16) {
   V1 = VLOAD(I,x:x+31,y); 
   V2 = VLOAD(W,0:7); 
   V3 = VMUL(V2, SELECT(V2, {j}), 
             V1, SELECT(V1, {i+j}));
   V3 = VMAC(V3, V2, SELECT(V2,{j+2}),
                 V1, SELECT(V1,{i+j+2}}));
   V4 = VLOAD(I,x:x+31,y+1); 
   V3 = VMAC(V3, V2, SELECT(V4, {j+4}), 
                 V1, SELECT(V1, {i+j}));
   V3 = VMAC(V3, V2, SELECT(V4,{j+6}),
                 V1, SELECT(V1,{i+j+2}}));
   ....
   VSTORE(V2, O, x:x+15,y);
}   
\end{lstlisting}
\caption{A snippet of the generated 16-bit vector code for the running example in~\cref{fig:motivating-example}. 
VLOAD/VMUL/VMAC/VSTORE refers to vector load, vector multiplication, vector multiply-and-accumulate, and vector store. 
SELECT symbolically represents the data selection over a vector register for the i$^{th}$ row and j$^{th}$ column of 2D datapath multipliers.
}
    \label{fig:code-generated}
\end{figure}

\subsection{Auto-tuner} \label{subsec:auto-tuner}

Steps 1-5 in our multi-step compiler approach generates the vectorized code for a given specification of tensor convolution, a schedule from the auto-tuner, and workload sizes.
The auto-tuning capabilities of the Halide framework support only multi-staged pipelines~\cite{10.1145/2897824.2925952,10.1145/3306346.3322967}, but our focus is only on a single stage for the convolution.
Hence, we implemented custom auto-tuner in our approach exploring all possible schedules to find the best schedule for a given convolution specification and the workload sizes.
The search space of schedules include loop nest and data-layout optimizations.

\betterparagraph{Search space} The space of loop transformations include loop interchange, loop unroll and jamming, and the choice of loop for vectorization.
The space of data-layout optimizations include dimension permutation and data tiling.


\betterparagraph{Exploration} Our approach applies the  following pruning strategies: 1) unrolling of reduction loops to avoid memory traffic in writing and reading intermediate (partial) results, and 2) applying bounds on the unroll and jam factors to avoid code size explosion (AI Engine has only 16KB program memory) and also to avoid longer compilation times.
Our auto-tuner evaluates each point in the pruned search space by generating the vectorized C-code, compiling with the AI Engine compiler, and executing it using a cycle-accurate architecture simulator.
With performance as the primary optimization goal, our approach obtained a geometric mean performance improvement of 1.10$\times$ fewer cycles than the expert-written and tuned codes available for four workloads, showing that automatic exploration can find useful design points which are not obvious to humans.

\section{Experiments}
\label{sec:experiments}

We evaluated our approach over a total of 36 workloads involving a wide variety of operators and variations of CONV2D and CONV3D over two operand precisions (32-bit and 16-bit) on a single AI Engine.  Each workload represents a unique combination of a convolution operation, tensor shapes, and operand precision. 
The configuration is shown in Table \ref{tab:hw} and includes a 128KB local memory pre-loaded with all the data required for the evaluation of each workload. 
The configuration also includes a vector register file of size 256B (a total of 16 registers with each size as 128 bits) in between the SIMD datapath and the local memory. 
We used the AI Engine's cycle-accurate simulator to evaluate the functionality and performance of our generated codes.
We define the performance (MACs/Cycle) of an implementation of a tensor convolution as the total number of MAC operations in the convolution divided by the total number of execution cycles taken by the implementation.

\begin{table}[!ht]
\centering
\caption{The AI Engine configuration used in our evaluation.}
\begin{tabular}{|c|c|c|}
\hline
\textbf{Parameter}             & 32-bit & 16-bit        \\ \hline
\textbf{2D SIMD data path} & 8 x 1  & 16 x 2        \\ \hline
\textbf{Peak compute}                       & 8 MACs/cycle & 32 MACs/cycle \\ \hline
\textbf{Scratchpad memory}    & \multicolumn{2}{|c|}{128 KB @ 96B/cycle}         \\ \hline
\textbf{Scratchpad memory ports}    & \multicolumn{2}{|c|}{32B 2 read and 1 write}         \\ \hline
\textbf{Vector register file}  & \multicolumn{2}{|c|}{256 B}        \\ \hline
\end{tabular}
\label{tab:hw}

\end{table}

\subsection{CONV2D in Computer Vision}

In the following experiments, we compare two experimental variants: 1) Code written by an expert (for 3$\times$3 and 5$\times$5 filters) available as part of the Xilinx's AI Engine compiler infrastructure, 
2) Code generated by our approach leveraging the auto-tuner. 
Both codes are designed to produce a 256$\times$16 tile of a larger image.
We observe from~\cref{fig:Conv2D-2D-Variants} that our approach achieved a geometric mean performance improvement of 1.10$\times$ from the Halide codes compared with the available expert-written codes.


\begin{figure}[!ht]
    \centering
    \includegraphics[width=\linewidth]{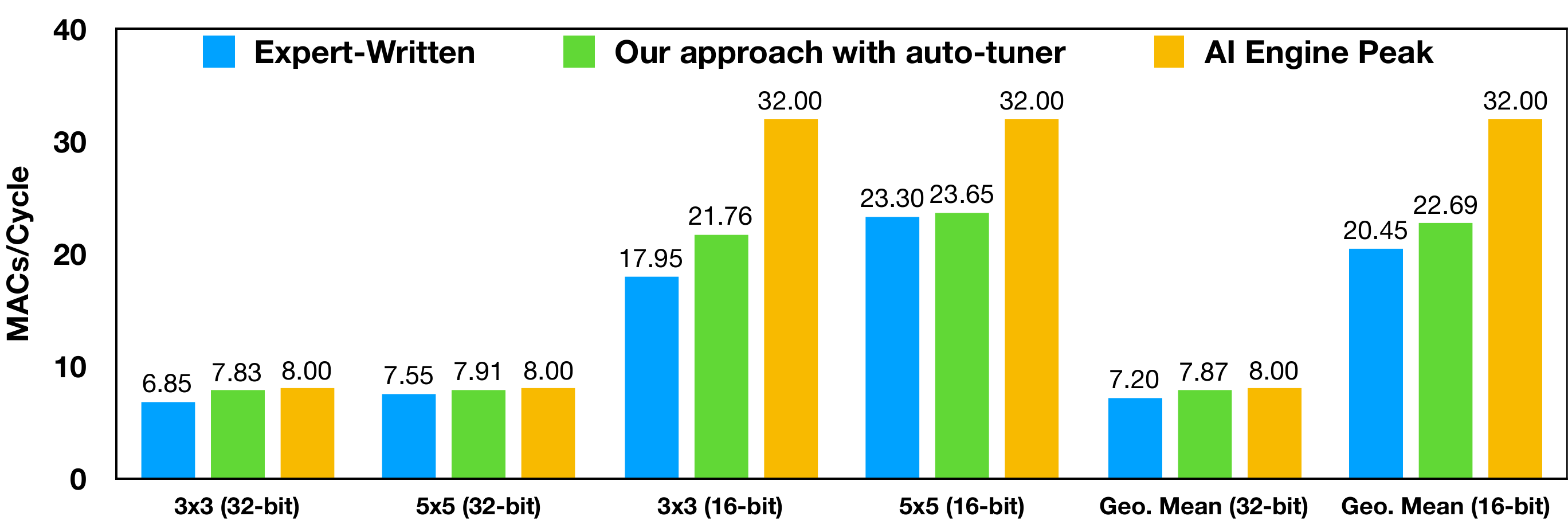}
    \caption{Comparison of our approach with auto-tuner against the available expert-written codes for CONV2D operation with 3$\times$3 and 5$\times$5 filters.}
    \label{fig:Conv2D-2D-Variants}
\end{figure}

The auto-tuner of our approach was able to find better schedules than used in the expert-written codes (roof-line graphs for the workloads is shown in~\cref{fig:Conv2D-2D-Variants-roofline}), including non-unit unroll and jam factors along the image height ({\tt loop-y}) dimension for better reuse.
These non-unit factors also enabled more opportunities in the loop body for the downstream compilers to perform better software pipelining.
Furthermore, since workload sizes are also expressed in the Halide codes, our approach annotated the loops of generated codes with pragmas about the loop sizes to help the downstream compilers, especially helping the automatic software pipelining to accurately estimate the pre-amble and post-amble set up overheads and generate better VLIW code.
Such overheads can be significant, particularly for tiled inner loops executed many times.


\begin{figure}[!ht]
    \centering
    \includegraphics[width=\linewidth]{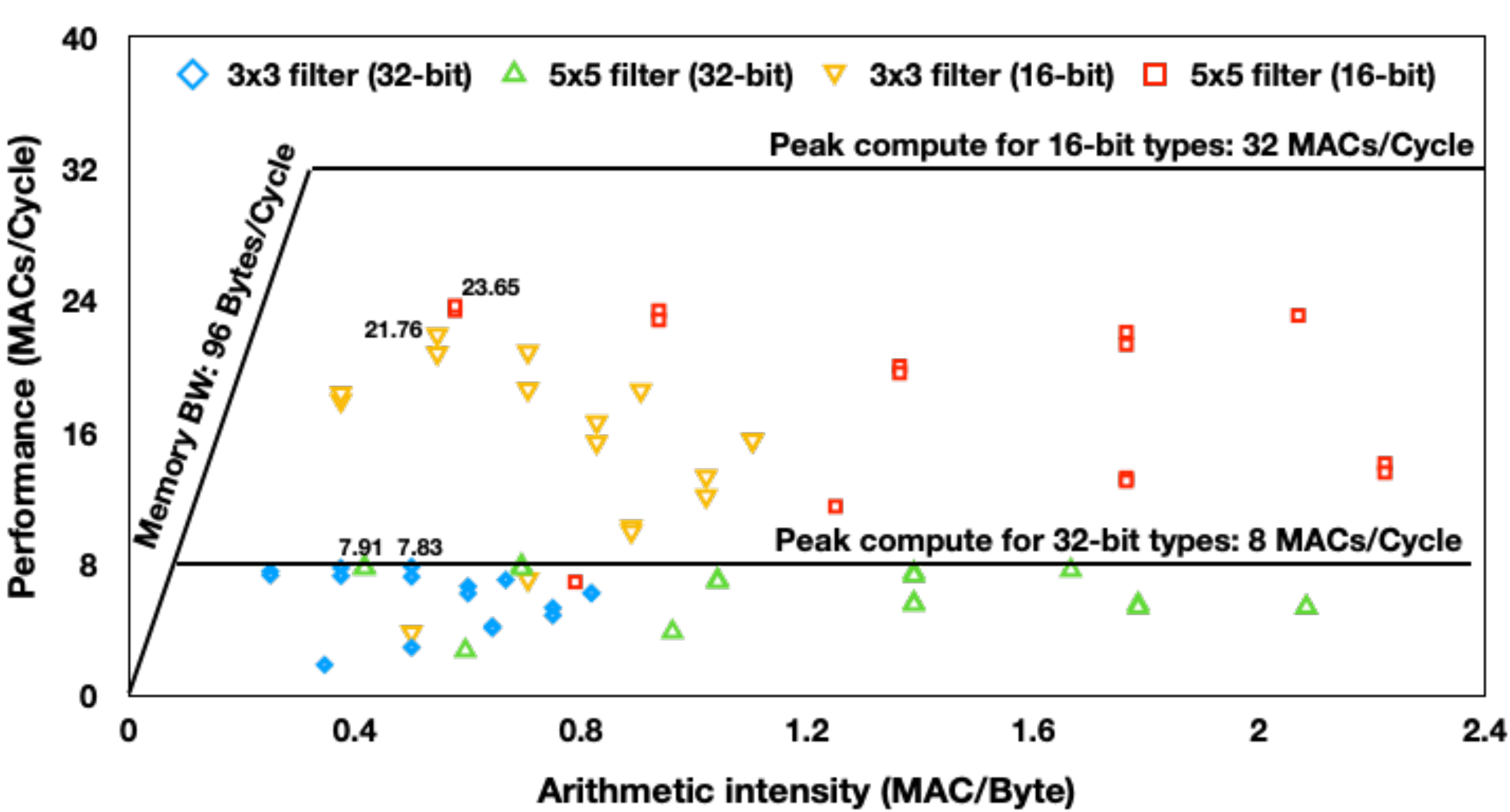}
    \caption{Roof-line graphs of four workloads considered in~\cref{fig:Conv2D-2D-Variants}, where each data point is a  schedule explored by the auto-tuner.}
    \label{fig:Conv2D-2D-Variants-roofline}
\end{figure}

In case of the 3x3 and 5x5 filters with 16-bit operands, the total number of fusible logical 1D vector multiplications corresponding to each row of the filters is an odd number.
Hence, our approach padded the filters with an additional column to generate even number of fusible 1D operations and map onto the two columns present in the 2D SIMD datapath for 16-bit types.
But, expert-written codes fused logical 1D vector multiplications corresponding to different rows of the filters, there by avoiding the padding.
This was accomplished by carefully merging the required input image data from different rows into a single vector register.
Our approach currently doesn't exploit this optimization strategy and would require additional analysis to enable it.
However, we see that the code generated using our approach is still able to perform better than the expert-written code by leveraging loop unroll and jam transformations.

\begin{table}[!ht]
\caption{CONV2D workloads of Computer Vision used in our evaluation and optimal schedules from auto-tuner}
\centering
\scalebox{1}{
\begin{tabular}{|c|c|c|c|c|c|c|c|c|}
\hline
\multirow{4}{*}{\textbf{\begin{tabular}[c]{@{}c@{}}Output \\ (O) size\end{tabular}}} & \multirow{4}{*}{\textbf{\begin{tabular}[c]{@{}c@{}}Weight \\ (W) size\end{tabular}}} & \multirow{4}{*}{\textbf{\begin{tabular}[c]{@{}c@{}}Input \\ (I) size\end{tabular}}} & \multirow{4}{*}{\textbf{\#MACs}} & \multicolumn{5}{c|}{\textbf{\begin{tabular}[c]{@{}c@{}}Optimal schedule \\ from auto-tuner\end{tabular}}}                                                                                             \\ \cline{5-9} 
                                                                                     &                                                                                      &                                                                                     &                                  & \multicolumn{4}{c|}{\textbf{\begin{tabular}[c]{@{}c@{}}Unroll and \\ Jam factors\end{tabular}}} & \multicolumn{1}{l|}{\multirow{3}{*}{\textbf{\begin{tabular}[c]{@{}l@{}}Loop\\ order\end{tabular}}}} \\ \cline{5-8}
                                                                                     &                                                                                      &                                                                                     &                                  & \multicolumn{2}{c|}{\textbf{32-bit}}            & \multicolumn{2}{c|}{\textbf{16-bit}}          & \multicolumn{1}{l|}{}                                                                               \\ \cline{5-8}
                                                                                     &                                                                                      &                                                                                     &                                  & \textbf{x}             & \textbf{y}             & \textbf{x}            & \textbf{y}            & \multicolumn{1}{l|}{}                                                                               \\ \hline
\multirow{10}{*}{\begin{tabular}[c]{@{}c@{}}256 \\ x 16\end{tabular}}                & 2 x 2                                                                                & 264 x 17                                                                            & 16384                            & 1                      & 4                      & 1                     & 8                     & xy                                                                                                  \\ \cline{2-9} 
                                                                                     & 3 x 3                                                                                & 264 x 18                                                                            & 36864                            & 1                      & 4                      & 1                     & 2                     & xy                                                                                                  \\ \cline{2-9} 
                                                                                     & 4 x 4                                                                                & 264 x 19                                                                            & 65536                            & 1                      & 2                      & 1                     & 1                     & xy                                                                                                  \\ \cline{2-9} 
                                                                                     & 5 x 5                                                                                & 264 x 20                                                                            & 102400                           & 1                      & 2                      & 1                     & 1                     & xy                                                                                                  \\ \cline{2-9} 
                                                                                     & 6 x 6                                                                                & 264 x 21                                                                            & 147456                           & 1                      & 1                      & 1                     & 1                     & xy                                                                                                  \\ \cline{2-9} 
                                                                                     & 7 x 7                                                                                & 264 x 22                                                                            & 200704                           & 1                      & 1                      & 1                     & 1                     & xy                                                                                                  \\ \cline{2-9} 
                                                                                     & 8 x 8                                                                                & 264 x 23                                                                            & 262144                           & 1                      & 4                      & 1                     & 1                     & xy                                                                                                  \\ \cline{2-9} 
                                                                                     & 9 x 9                                                                                & 264 x 24                                                                            & 331776                           & 1                      & 4                      & 1                     & 1                     & xy                                                                                                  \\ \cline{2-9} 
                                                                                     & 10 x 10                                                                              & 264 x 25                                                                            & 409600                           & 1                      & 4                      & 1                     & 4                     & xy                                                                                                  \\ \cline{2-9} 
                                                                                     & 11 x 11                                                                              & 264 x 26                                                                            & 495616                           & 1                      & 4                      & 1                     & 4                     & xy                                                                                                  \\ \hline
\end{tabular}
}
\label{tab:2DCONV-filters}
\end{table}

\begin{figure}[!ht]
    \centering
    \includegraphics[width=\linewidth]{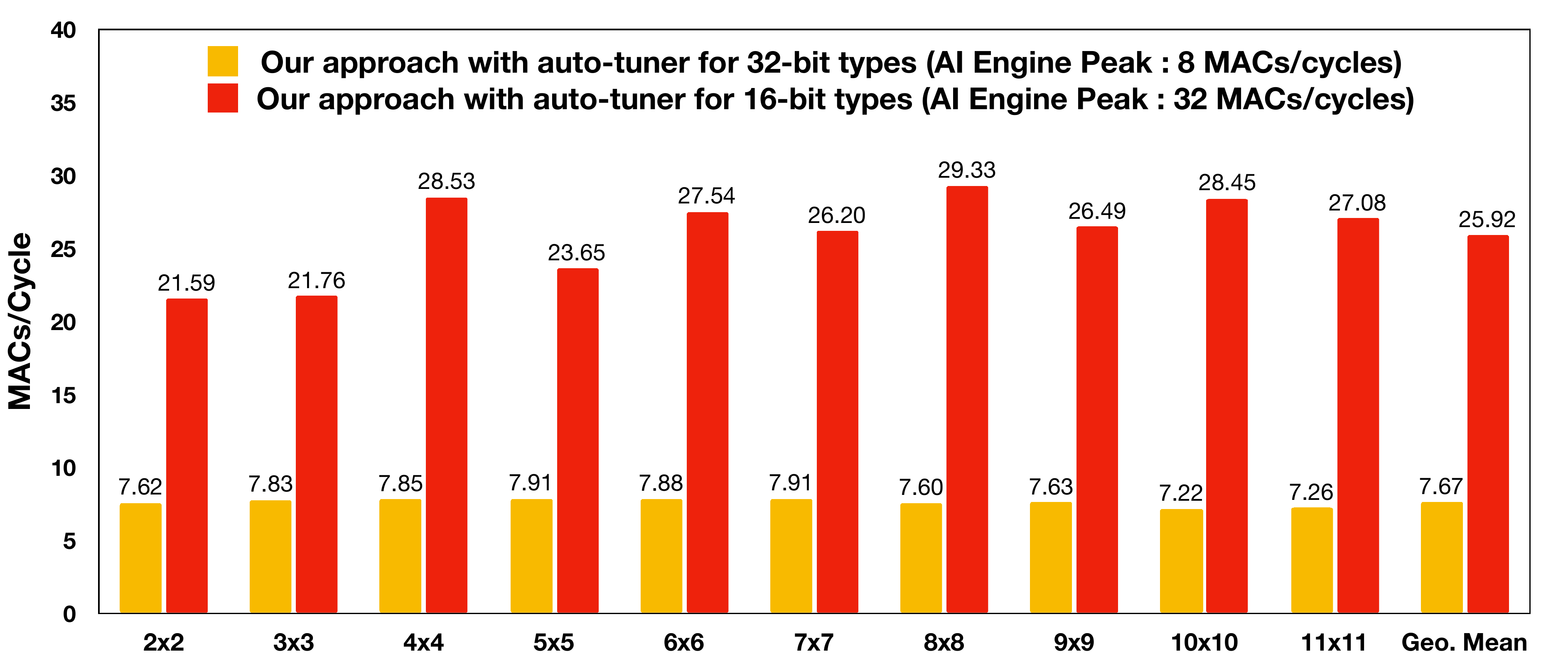}
    \caption{Performance of our approach generated codes for CONV2D workloads of Computer Vision over filter sizes from 2 to 11.}
    \label{fig:Conv2D-2D-Filters}
\end{figure}

In addition to the 3x3 and 5x5 filters, we have evaluated other filter sizes commonly used in Computer Vision applications. Table~\ref{tab:2DCONV-filters} presents those workload sizes, total MAC operations involved in each workload, and optimal schedules reported by the auto-tuner. 
We padded each non-even sized 16-bit filter with an additional column for evaluation, but we used the MACs obtained by the filter without padding while computing the performance (MACs/cycle). 
As can be observed from~\cref{fig:Conv2D-2D-Filters}, our approach achieved a geometric mean performance of 7.67 and 25.92 MACs/cycle for 32-bit and 16-bit types respectively for the workloads in Table~\ref{tab:2DCONV-filters}.
The auto-tuner chose the {\tt loop-x} for vectorization for all the workloads, because it has more reuse opportunities and has larger number of iterations compared to the {\tt loop-y}.
The optimal unroll and jam factors are not the same for all the workloads and also vary for different precisions of the same filter size.
Even though increasing unroll and jam factors improve the reuse opportunities, but it often resulted in register spills after a threshold and also interfered with software pipelining of inner loops.
Furthermore, larger unroll and jam factors along the {\tt loop-x} resulted in larger connected components of the reuse graph and required larger vector register than the maximum possible (e.g., 1024b for 32-bit operands) in the hardware. 

\subsection{CONV2D in Deep Learning}

\begin{table*}[!ht]
\centering
\caption{CONV2D workloads of deep learning used in our evaluation (variable names described in~\cref{sec:background}) and optimal schedules.
}
\scalebox{0.9}{
\begin{tabular}{|c|c|c|c|c|c|ccccccccc}
\hline
\multirow{3}{*}{\textbf{\begin{tabular}[c]{@{}c@{}}CONV\\ type\end{tabular}}} & \multirow{3}{*}{\textbf{\begin{tabular}[c]{@{}c@{}}Output \\ (O) size\\ (XxYxK)\end{tabular}}} & \multirow{3}{*}{\textbf{\begin{tabular}[c]{@{}c@{}}Filter \\ (F) size\\ (RxSxCxK)\end{tabular}}} & \multirow{3}{*}{\textbf{\begin{tabular}[c]{@{}c@{}}Input\\ (I) size\\ (X'xY'xC)\end{tabular}}} & \multirow{3}{*}{\textbf{\#MACs}} & \multirow{3}{*}{\textbf{Precision}} & \multicolumn{9}{c}{\textbf{Optimal schedules from the auto-tuner}}                                                                                                                                                                                                                                                                                                                                                                                                                                                                   \\ \cline{7-15} 
                                                                              &                                                                                                &                                                                                                  &                                                                                              &                                  &                                     & \multicolumn{3}{c|}{\textbf{Data layouts}}                                                                    & \multicolumn{1}{c|}{\multirow{2}{*}{\textbf{\begin{tabular}[c]{@{}c@{}}Vector\\ loop\end{tabular}}}} & \multicolumn{1}{c|}{\multirow{2}{*}{\textbf{\begin{tabular}[c]{@{}c@{}}SW\\ loop\end{tabular}}}} & \multicolumn{3}{c|}{\textbf{\begin{tabular}[c]{@{}c@{}}Unroll and \\ Jam factors\end{tabular}}}     & \multicolumn{1}{c|}{\multirow{2}{*}{\textbf{\begin{tabular}[c]{@{}c@{}}Loop \\ order\end{tabular}}}} \\ \cline{7-9} \cline{12-14}
                                                                              &                                                                                                &                                                                                                  &                                                                                              &                                  &                                     & \multicolumn{1}{c|}{\textbf{O}} & \multicolumn{1}{c|}{\textbf{W}}           & \multicolumn{1}{c|}{\textbf{I}} & \multicolumn{1}{c|}{}                                                                                & \multicolumn{1}{c|}{}                                                                            & \multicolumn{1}{c|}{\textbf{x}} & \multicolumn{1}{c|}{\textbf{y}} & \multicolumn{1}{c|}{\textbf{k}} & \multicolumn{1}{c|}{}                                                                                \\ \hline
\multirow{6}{*}{\textbf{(REG)}}                                               & \multirow{14}{*}{128x2x16}                                                                     & \multirow{2}{*}{3x3x8x16}                                                                        & \multirow{2}{*}{144x4x8}                                                                     & \multirow{2}{*}{294912}          & 32-bit                              & \multicolumn{1}{c|}{XYK}        & \multicolumn{1}{c|}{(K/8)(C/8)SR(8)(8)}   & \multicolumn{1}{c|}{(C/8)Y'X'(8)} & \multicolumn{1}{c|}{k}                                                                               & \multicolumn{1}{c|}{x}                                                                           & \multicolumn{1}{c|}{1}          & \multicolumn{1}{c|}{2}          & \multicolumn{1}{c|}{1}          & \multicolumn{1}{c|}{kyx}                                                                             \\ \cline{6-15} 
                                                                              &                                                                                                &                                                                                                  &                                                                                              &                                  & 16-bit                              & \multicolumn{1}{c|}{KYX}        & \multicolumn{1}{c|}{K(C/2)SR(2)}          & \multicolumn{1}{c|}{(C/2)Y'X'(2)} & \multicolumn{1}{c|}{x}                                                                               & \multicolumn{1}{c|}{x}                                                                           & \multicolumn{1}{c|}{1}          & \multicolumn{1}{c|}{1}          & \multicolumn{1}{c|}{1}          & \multicolumn{1}{c|}{yxk}                                                                             \\ \cline{3-15} 
                                                                              &                                                                                                & \multirow{2}{*}{5x5x8x16}                                                                        & \multirow{2}{*}{144x6x8}                                                                     & \multirow{2}{*}{819200}          & 32-bit                              & \multicolumn{1}{c|}{KYX}        & \multicolumn{1}{c|}{KCSR}                 & \multicolumn{1}{c|}{CY'X'}        & \multicolumn{1}{c|}{x}                                                                               & \multicolumn{1}{c|}{x}                                                                           & \multicolumn{1}{c|}{1}          & \multicolumn{1}{c|}{1}          & \multicolumn{1}{c|}{1}          & \multicolumn{1}{c|}{kyx}                                                                             \\ \cline{6-15} 
                                                                              &                                                                                                &                                                                                                  &                                                                                              &                                  & 16-bit                              & \multicolumn{1}{c|}{KYX}        & \multicolumn{1}{c|}{K(C/2)SR(2)}          & \multicolumn{1}{c|}{(C/2)Y'X'(2)} & \multicolumn{1}{c|}{x}                                                                               & \multicolumn{1}{c|}{x}                                                                           & \multicolumn{1}{c|}{1}          & \multicolumn{1}{c|}{2}          & \multicolumn{1}{c|}{1}          & \multicolumn{1}{c|}{kyx}                                                                             \\ \cline{3-15} 
                                                                              &                                                                                                & \multirow{2}{*}{7x7x8x16}                                                                        & \multirow{2}{*}{144x8x8}                                                                     & \multirow{2}{*}{1605632}         & 32-bit                              & \multicolumn{1}{c|}{KYX}        & \multicolumn{1}{c|}{KCSR}                 & \multicolumn{1}{c|}{CY'X'}        & \multicolumn{1}{c|}{x}                                                                               & \multicolumn{1}{c|}{x}                                                                           & \multicolumn{1}{c|}{1}          & \multicolumn{1}{c|}{2}          & \multicolumn{1}{c|}{1}          & \multicolumn{1}{c|}{kyx}                                                                             \\ \cline{6-15} 
                                                                              &                                                                                                &                                                                                                  &                                                                                              &                                  & 16-bit                              & \multicolumn{1}{c|}{KYX}        & \multicolumn{1}{c|}{K(C/2)SR(2)}          & \multicolumn{1}{c|}{(C/2)Y'X'(2)} & \multicolumn{1}{c|}{x}                                                                               & \multicolumn{1}{c|}{x}                                                                           & \multicolumn{1}{c|}{1}          & \multicolumn{1}{c|}{2}          & \multicolumn{1}{c|}{1}          & \multicolumn{1}{c|}{kyx}                                                                             \\ \cline{1-1} \cline{3-15} 
\multirow{2}{*}{\textbf{(PW)}}                                                &                                                                                                & \multirow{2}{*}{1x1x8x16}                                                                        & \multirow{2}{*}{144x2x8}                                                                     & \multirow{2}{*}{32768}           & 32-bit                              & \multicolumn{1}{c|}{XYK}        & \multicolumn{1}{c|}{(K/8)(C/8)SR(8)(8)}   & \multicolumn{1}{c|}{(C/8)Y'X'(8)} & \multicolumn{1}{c|}{k}                                                                               & \multicolumn{1}{c|}{x}                                                                           & \multicolumn{1}{c|}{1}          & \multicolumn{1}{c|}{2}          & \multicolumn{1}{c|}{1}          & \multicolumn{1}{c|}{kyx}                                                                             \\ \cline{6-15} 
                                                                              &                                                                                                &                                                                                                  &                                                                                              &                                  & 16-bit                              & \multicolumn{1}{c|}{YXK}        & \multicolumn{1}{c|}{(K/16)SR(C/2)(16)(2)} & \multicolumn{1}{c|}{Y'X'C}        & \multicolumn{1}{c|}{k}                                                                               & \multicolumn{1}{c|}{k}                                                                           & \multicolumn{1}{c|}{1}          & \multicolumn{1}{c|}{2}          & \multicolumn{1}{c|}{1}          & \multicolumn{1}{c|}{xyk}                                                                             \\ \cline{1-1} \cline{3-15} 
\multirow{4}{*}{\textbf{(SS)}}                                                &                                                                                                & \multirow{2}{*}{1x3x8x16}                                                                        & \multirow{2}{*}{144x4x8}                                                                     & \multirow{2}{*}{98304}           & 32-bit                              & \multicolumn{1}{c|}{XYK}        & \multicolumn{1}{c|}{(K/8)(C/8)SR(8)(8)}   & \multicolumn{1}{c|}{(C/8)Y'X'(8)} & \multicolumn{1}{c|}{k}                                                                               & \multicolumn{1}{c|}{p}                                                                           & \multicolumn{1}{c|}{1}          & \multicolumn{1}{c|}{2}          & \multicolumn{1}{c|}{1}          & \multicolumn{1}{c|}{kyx}                                                                             \\ \cline{6-15} 
                                                                              &                                                                                                &                                                                                                  &                                                                                              &                                  & 16-bit                              & \multicolumn{1}{c|}{KYX}        & \multicolumn{1}{c|}{K(C/2)SR(2)}          & \multicolumn{1}{c|}{(C/2)Y'X'(2)} & \multicolumn{1}{c|}{x}                                                                               & \multicolumn{1}{c|}{x}                                                                           & \multicolumn{1}{c|}{1}          & \multicolumn{1}{c|}{2}          & \multicolumn{1}{c|}{1}          & \multicolumn{1}{c|}{kyx}                                                                             \\ \cline{3-15} 
                                                                              &                                                                                                & \multirow{2}{*}{3x1x8x16}                                                                        & \multirow{2}{*}{144x2x8}                                                                     & \multirow{2}{*}{98304}           & 32-bit                              & \multicolumn{1}{c|}{XYK}        & \multicolumn{1}{c|}{(K/8)(C/8)SR(8)(8)}   & \multicolumn{1}{c|}{(C/8)Y'X'(8)} & \multicolumn{1}{c|}{k}                                                                               & \multicolumn{1}{c|}{x}                                                                           & \multicolumn{1}{c|}{1}          & \multicolumn{1}{c|}{1}          & \multicolumn{1}{c|}{1}          & \multicolumn{1}{c|}{kyx}                                                                             \\ \cline{6-15} 
                                                                              &                                                                                                &                                                                                                  &                                                                                              &                                  & 16-bit                              & \multicolumn{1}{c|}{YXK}        & \multicolumn{1}{c|}{(K/16)SR(C/2)(16)(2)} & \multicolumn{1}{c|}{Y'X'C}        & \multicolumn{1}{c|}{k}                                                                               & \multicolumn{1}{c|}{k}                                                                           & \multicolumn{1}{c|}{1}          & \multicolumn{1}{c|}{2}          & \multicolumn{1}{c|}{1}          & \multicolumn{1}{c|}{xyk}                                                                             \\ \cline{1-1} \cline{3-15} 
\multirow{2}{*}{\textbf{(DS)}}                                                &                                                                                                & \multirow{2}{*}{3x3x16x16}                                                                       & \multirow{2}{*}{144x4x16}                                                                    & \multirow{2}{*}{36864}           & 32-bit                              & \multicolumn{1}{c|}{KYX}        & \multicolumn{1}{c|}{KCSR}                 & \multicolumn{1}{c|}{CY'X'}        & \multicolumn{1}{c|}{x}                                                                               & \multicolumn{1}{c|}{x}                                                                           & \multicolumn{1}{c|}{1}          & \multicolumn{1}{c|}{2}          & \multicolumn{1}{c|}{1}          & \multicolumn{1}{c|}{kyx}                                                                             \\ \cline{6-15} 
                                                                              &                                                                                                &                                                                                                  &                                                                                              &                                  & 16-bit                              & \multicolumn{1}{c|}{KYX}        & \multicolumn{1}{c|}{KCSR}                 & \multicolumn{1}{c|}{CY'X'}        & \multicolumn{1}{c|}{x}                                                                               & \multicolumn{1}{c|}{x}                                                                           & \multicolumn{1}{c|}{1}          & \multicolumn{1}{c|}{2}          & \multicolumn{1}{c|}{1}          & \multicolumn{1}{c|}{kyx}                                                                             \\ \hline
\multirow{2}{*}{\textbf{(FC)}}                                                & \multirow{2}{*}{4096x1x1}                                                                      & \multirow{2}{*}{1x1x8x4096}                                                                      & \multirow{2}{*}{16x1x8}                                                                      & \multirow{2}{*}{32768}           & 32-bit                              & \multicolumn{1}{c|}{XYK}        & \multicolumn{1}{c|}{(K/8)(C/8)SR(8)(8)}   & \multicolumn{1}{c|}{(C/8)Y'X'(8)} & \multicolumn{1}{c|}{k}                                                                               & \multicolumn{1}{c|}{k}                                                                           & \multicolumn{1}{c|}{1}          & \multicolumn{1}{c|}{1}          & \multicolumn{1}{c|}{1}          & \multicolumn{1}{c|}{kyx}                                                                             \\ \cline{6-15} 
                                                                              &                                                                                                &                                                                                                  &                                                                                              &                                  & 16-bit                              & \multicolumn{1}{c|}{YXK}        & \multicolumn{1}{c|}{(K/16)SR(C/2)(16)(2)} & \multicolumn{1}{c|}{Y'X'C}        & \multicolumn{1}{c|}{k}                                                                               & \multicolumn{1}{c|}{k}                                                                           & \multicolumn{1}{c|}{1}          & \multicolumn{1}{c|}{1}          & \multicolumn{1}{c|}{1}          & \multicolumn{1}{c|}{xyk}                                                                             \\ \hline
\end{tabular}
}
\label{tab:conv2d-3d-images}
\end{table*}

We considered a wide variety of CONV2D operations in the deep learning domain such as regular (REG) CONV2D over various filter sizes, point-wise (PW), spatially separable (SS), depth-wise separable (DS), and fully-connected (FC) operations.
Table~\ref{tab:conv2d-3d-images} presents those workload sizes (with unit batch size, i.e., N = 1), total MAC operations involved in each workload, and optimal schedules reported by the auto-tuner. 
Since the memory footprint of typical CONV2D operations don't fit into the local memory, we chose the similar output and input tensor memory footprint used in Table~\ref{tab:2DCONV-filters}.
As can be observed from~\cref{fig:Conv2D-3D-Filters}, our approach achieved a geometric mean performance of 7.67 and 22.53 MACs/cycle for 32-bit and 16-bit types respectively for the workloads in Table~\ref{tab:conv2d-3d-images}.


\begin{figure}[!ht]
    \centering
    \includegraphics[width=\linewidth]{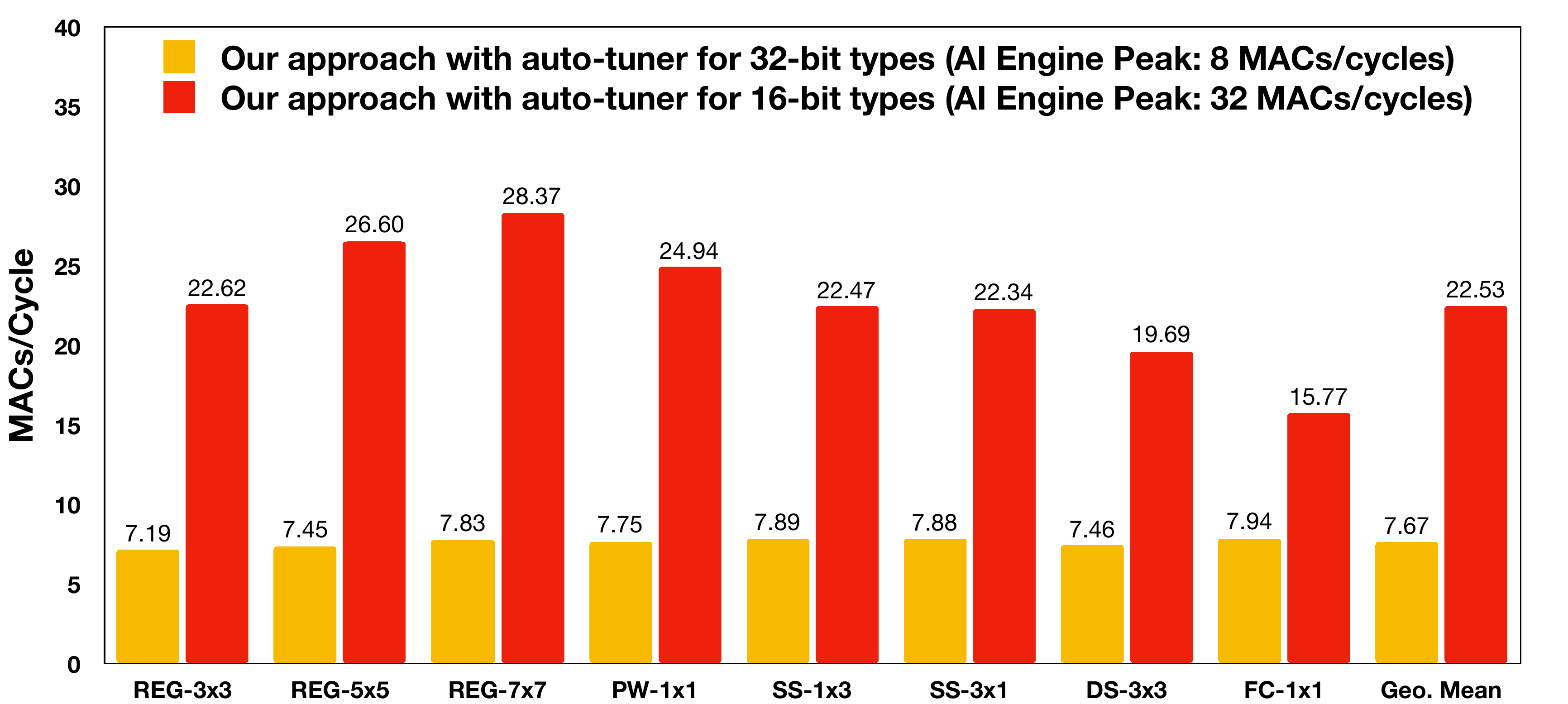}
    \caption{Performance of our approach generated codes for CONV2D workloads (shown in Table~\ref{tab:conv2d-3d-images}) of Deep Learning.}
    \label{fig:Conv2D-3D-Filters}
\end{figure}

The auto-tuner chose either {\tt loop-x} or {\tt loop-k} for vectorizing the workloads, because the number of iterations of remaining loops are smaller than the vector length.
The auto-tuner identified the vectorization along the {\tt loop-x} to be beneficial for the REG-5x5, REG-7x7 workloads, because there exists more opportunities for vector register reuse (convolutional reuse) along the {\tt loop-x} with the larger kernels sizes.
But, for the workloads such as PW (REG-1x1), FC that have either less or no convolutional reuse along the {\tt loop-x}, the vectorization was performed on the {\tt loop-k}.

\begin{figure}[!ht]
    \centering
    \includegraphics[width=\linewidth]{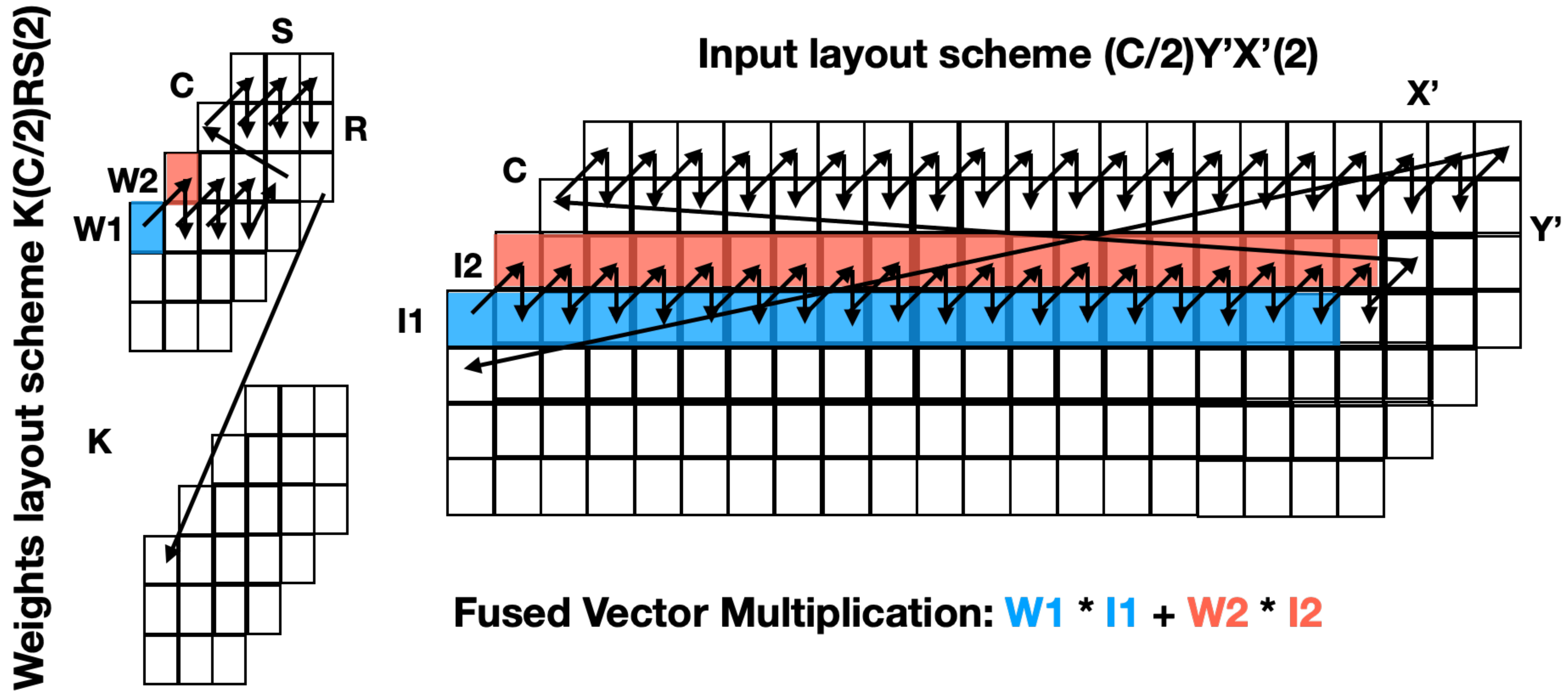}
    \caption{Data-layouts of input and weight tensors of the 16-bit REG-3x3 workload (Table~\ref{tab:conv2d-3d-images}), to enable the fusion of 1D logical vector multiplications along the channels, there by avoiding the padding required for weights.}
    \label{fig:data-tiling-example}
\end{figure}

In these workloads, there exists an even number of fusible logical 1D vector multiplications corresponding to the filter channels, hence our approach didn't require any padding to the filter tensors (unlike in Table~\ref{tab:2DCONV-filters}), except for the depth-wise CONV2D workload which has only one channel.
However, the data-layouts of these workload tensors need to be modified to support the fusion of 1D logical vector multiplications along the channels.
An example data-layout for the input and weights of the 16-bit REG-3x3 workload for the fusion along channels is shown in~\cref{fig:data-tiling-example}, where the data-layout scheme for the input tensor (C/2)Y'X'(2) refers to first laying out a block of two channels followed by width, height, and remaining channels.

Along with the advantages of avoiding padding, data-layouts can be used for exploring better schedules as well.
Such data-layout schemes over the workload tensors should respect two constraints: 1) The required number of data elements of each operand of the fused vector multiplication should fit into the maximum vector register size (e.g., 32 unique 16-bit input data elements for the vector multiplication in~\cref{fig:data-tiling-example} can fit into a 1024b vector register which is the maximum), and 2) The required data selection parameters over the vector register should respect the shuffle network constraints.
Our auto-tuner was able to automatically explore a variety of such valid data-layout schemes in our evaluation.  Although the resulting data-layouts can be implemented by the architecture, they can be rather complex and non-intuitive (e.g., (K/16)SR(C/2)(16)(2) in Table~\ref{tab:conv2d-3d-images}).  Manually identifying such a data layout and writing the corresponding instrisic-based code is extremely challenging and error-prone, even for experts, thereby demonstrating the benefits of our automatic approach.

In Table~\ref{tab:conv2d-3d-images}, we see that the arithmetic intensity of the FC workload for the 16-bit is to the left-side of the inflection point of the roof-line graph of the AI engine, indicating memory-bound execution.  
This is expected, since the FC workload has little opportunity for data reuse within a single convolution operation.
The workload peak performance based on its arithmetic intensity is 21.22 MACs/cycle, and our approach achieved 15.77 MACs/cycle or 75\% of the workload peak.

\subsection{CONV3D}


In this evaluation, we focused on the simpler CONV3D workloads to further demonstrate the applicability of our approach. 
The output sizes in these workloads are the same as in the CONV2D workloads in Table~\ref{tab:conv2d-3d-images}, i.e., 168x2x16, and the weight tensor sizes are 3x3x3, 5x5x5, and 7x7x7 which are popular in the 3D CNN models~\cite{ahmed2018survey,hou2017endtoend}.
Since the number of fusible 1D logical vector multiplications corresponding to any dimension of the weight tensor in these workloads are odd, we have padded the weights with an additional column for each row. 
With this padding, our approach achieved a geometric mean performance of 7.55 and 21.60 MACs/cycle for 32-bit and 16-bit types, respectively shown in~\cref{fig:Conv3D-3D-Filters}.

\begin{figure}[!ht]
    \centering
    \includegraphics[width=\linewidth]{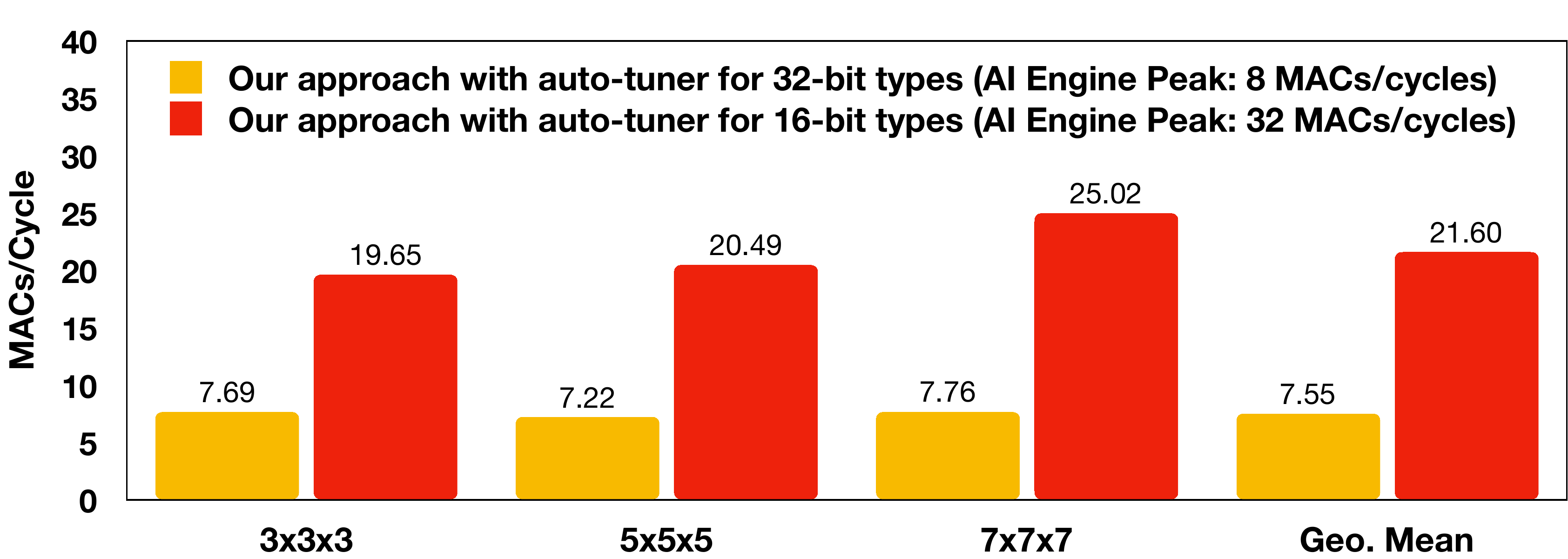}
    \caption{Performance of our approach generated codes for CONV3D workloads with weight sizes as 3x3x3, 5x5x5, 7x7x7.}
    \label{fig:Conv3D-3D-Filters}
\end{figure}

Overall, our evaluation over all the workloads shows geometric means of 7.6 and 23.3 MACs/cycle for 32-bit and 16-bit operands (which represent 95.9\% and 72.8\% of the peak performance respectively).
This difference in efficiency is not surprising, since it is more challenging to utilize two columns in the SIMD data path in the case of 16-bit operands, compared to a single column in the case of 32-bit operands.  However, the absolute performance in the 16-bit case is still significantly higher than the 32-bit case, despite a lower efficiency.
\section{Related Work}
\label{sec:related-work}

High-level and domain-specific compiler frameworks~\cite{ragan2013halide,polymage,tiramisu,tc,tvm,polymage-fpga,diesel,t2s-tensor} have been shown to improve the productivity of application programmers, while generating high-performance code for a variety of architectures including CPUs, GPUs, FPGAs, Spatial accelerators, and distributed systems.
Notably, the Halide framework~\cite{ragan2013halide} for image processing pipelines has gained popular attention in the academic and industrial world. 
Recently, Vocke et al.~\cite{10.1145/3106343} extended the Halide framework to support specialized Digital Signal Processors (DSPs), mainly focusing on SIMD instruction sets and heterogeneous scratchpad memories of the Intel Imaging Processing Units (IPUs). 
Furthermore, Halide has the support for the Hexagon Vector eXtensions (HVX) on the Qualcomm Hexagon DSP processors.
However, none of the above prior work focused on targeting the 2D SIMD datapaths and the shuffle interconnection networks, which are unique to the AI Engine.

To the best of our knowledge, the only prior work on auto-vectorizing for a 2D SIMD datapath is the work by Dasika et al.~\cite{6113792}, where the authors have proposed a greedy compiler approach implemented as an extension to Trimaran~\cite{trimiran} compiler, to identify a sequence of back to back vector operations for execution on their PEPSC's architecture chained FPUs. 
But, our approach identifies a group of such back to back dependent (i.e., fusible 1D logical) vector operations by searching in the triplet representation, a simplified and symbolic view of the convolution loop body.

Exploiting vector register reuse (including partial reuse) on SIMD units is a vital optimization to achieve high-performance, and prior work exploited the reuse by shuffling the vector registers using the data manipulation/shuffle units~\cite{stock2014vectorization,partial,6332297}. 
But, our approach constructs a larger vector load covering the loads having reuse and uses the AI Engine's unique shuffle network to select the desired elements.  
Furthermore, our approach uses the shuffle network to address the unaligned vector loads and scalar broadcasts without requiring any additional hardware support.

The vector codes generated by our approach are viewed as high-performance primitives that are intended to execute on a single AI Engine.
These primitives are composed and integrated by a high-level compiler to run larger tensor convolutions across multiple AI Engines.
Some of the prior works that have followed the similar strategy of automating the library/primitive development for the performance-critical kernels are SPIRAL~\cite{4536398} for the domain of linear transforms, ATLAS~\cite{Whaley01automatedempirical} for the basic linear algebra subroutines (BLAS), and FFTW~\cite{1386650} for the discrete Fourier transforms.

\section{Conclusions \& Future Work}
\label{sec:conclusion}

In this work, we introduced \ourtool{}{}, a high-level programming system built on the Halide framework, to generate high-performance vector codes for the tensor convolutions onto the Xilinx Versal AI Engine.
Our proposed multi-step compiler approach leverages the AI Engine’s unique capabilities of the 2D SIMD datapath and the shuffle interconnection networks to achieve close to the peak performance for various workloads.
Manually identifying best schedules and writing the corresponding intrinsic-based code is extremely challenging and error-prone, even for experts, thereby demonstrating the benefits of our automatic approach. 
Our results show geometric means of 7.6 and 23.3 MACs/cycle for 32-bit and 16-bit operands (which represent 95.9\% and 72.8\% of the peak performance respectively). For four of these workloads for which expert-written implementations were available to us, Vyasa achieved a geometric mean performance improvement of 1.10$\times$ from Halide code that is around 50$\times$ smaller than the expert-written C/C++ code.
In the future, we plan to extend our system to other computationally expensive linear algebra kernels.
Also, we plan to integrate the generated high-performance codes into a high-level compiler to run larger tensor convolutions across multiple AI Engines.

 \bibliographystyle{IEEEtran}
\bibliography{References}

\end{document}